\documentclass{aa}  

\usepackage{graphicx}
\usepackage{txfonts}
\usepackage{natbib}
\usepackage{tablefootnote}
\usepackage{multirow}
\usepackage{longtable}
\usepackage[colorlinks=true,allcolors=blue]{hyperref}
\newcommand{\orcid}[1]{\unskip\protect\href{https://orcid.org/#1}{\protect\includegraphics[width=8pt,clip]{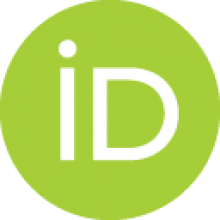}}}
\begin{document}

   \title{The phase curve of the ultra-hot Jupiter WASP-167b as seen by TESS}
    \titlerunning{The phase curve of WASP-167b}
    \authorrunning{Sz. K\'alm\'an et al.}

   \author{Sz.~K\'alm\'an
          \inst{1,2,3}\orcid{0000-0003-3754-7889}
          \and
          A.~Derekas\inst{4,5,7}\orcid{0000-0002-6526-9444}
          \and 
          Sz.~Csizmadia\inst{6}\orcid{0000-0001-6803-9698}
          \and 
          A.~P\'al\inst{1}\orcid{0000-0001-5449-2467}
          \and 
          R.~Szab\'o\inst{1,8}\orcid{0000-0002-3258-1909}
          \and 
          A.~M.~S.~Smith\inst{6}\orcid{0000-0002-2386-4341}
          \and
          K.~Nagy\inst{9}
          \and
          V.~Heged\H{u}s \inst{3, 5}\orcid{0000-0001-7699-1902}
          \and
          T.~Mitnyan\inst{7,10} \orcid{0000-0001-5803-3938}
          \and
          L. Szigeti\inst{4}
          \and
          Gy.~M. Szab\'o\inst{2,4,5}\orcid{0000-0002-0606-7930}
    }
\institute{ 
	Konkoly Observatory, Research Centre for Astronomy and Earth Sciences, HUN-REN, MTA Centre of Excellence, Konkoly-Thege Miklós út 15–17., H-1121, Hungary 
\and  
	HUN-REN-ELTE Exoplanet Research Group, Szombathely, Szent Imre h. u. 112., H-9700, Hungary       
\and
	ELTE E{\"o}tv{\"o}s Lor\'and University, Doctoral School of Physics,  Budapest, Pázmány Péter sétány 1/A, H-1117, Hungary
\and
	ELTE E{\"o}tv{\"o}s Lor\'and University, Gothard Astrophysical Observatory, Szombathely, Szent Imre h. u. 112., H-9700, Hungary
\and    
	MTA-ELTE  Lend{\"u}let "Momentum" Milky Way Research Group, Szombathely, Szent Imre h. u. 112., H-9700, Hungary
\and
        Deutsches Zentrum für Luft- und Raumfahrt, Institute of Planetary Research, Rutherfordstrasse 2, D-12489 Berlin, Germany
\and 
    HUN-REN-SZTE Stellar Astrophysics Research Group, H-6500 Baja, Szegedi út, Kt. 766, Hungary           
\and
    ELTE E\"otv\"os Lor\'and University, Institute of Physics, P\'azm\'any P\'eter s\'et\'any 1/A, H-1117 Budapest, Hungary
\and 
    University of Szeged, Doctoral School of Physics,  D\'om tér 9, 6720 Szeged, Hungary
\and 
        Baja Astronomical Observatory of University of Szeged, H-6500 Baja, Szegedi út, Kt. 766, Hungary
}

   \date{Received ...; accepted ...}

\abstract
   {Ultra-hot Jupiters (UHJs) orbiting pulsating A/F stars represent an important subset of the exoplanetary demographic, as they are excellent candidates for the study of exoplanetary atmospheres, as well as being astrophysical laboratories for the investigation of planet-to-star interactions.}
   {We analyse the \texttt{TESS} (Transiting Exoplanet Survey Satellite) light curve of the WASP-167 system, consisting of an F1V star and a substellar companion on a $\sim 2.02$ day orbit.}
   {We model the combination of the ellipsoidal variability and the Doppler beaming to measure the mass of WASP-167b, and the reflection effect to obtain constraints on the geometric albedo, while placing a special emphasis on noise separation. We implement a basic model to determine the dayside ($T_{\rm Day}$), nightside ($T_{\rm Night}$) and intrinsic ($T_{\rm Internal}$) temperatures of WASP-167b and put a constraint on its Bond albedo.}
   {We confirm the transit parameters of the planet seen in the literature. We find that a resonant $\sim 2P^{-1}$ stellar signal (which may originate from planet-to-star interactions) interferes with the phase curve analysis. After considerate treatment of this signal, we find $M_p = 0.34 \pm 0.22$~$M_J$. We measure a dayside temperature of $2790 \pm 100$ K, classifying WASP-167b as an UHJ. We find a $2\sigma$ upper limit of $0.51$ on its Bond albedo, and determine the geometric albedo at $0.34 \pm 0.11$ ($1 \sigma$ uncertainty).}
    {With an occultation depth of $106.8 \pm 27.3$ ppm in the \texttt{TESS} passband, the UHJ WASP-167b will be an excellent target for atmospheric studies, especially those at thermal wavelength ranges, where the stellar pulsations are expected to be be less influential.}
    
    \maketitle
\keywords{Planets and satellites: individual: WASP-167b -- Planets and satellites: atmospheres -- Planets and satellites: interior -- Planets and satellites: general -- Techniques: photometric}

 \section{Introduction}

Close-in, substellar mass companions orbiting pulsating host stars of A/F spectral types, such as WASP-33b \citep{2006MNRAS.372.1117C, 2010MNRAS.407..507C, 2011A&A...526L..10H}, HAT-P-2b\break \citep{2007ApJ...670..826B,2017ApJ...836L..17D}, HD 31221b \citep{2023arXiv230504000K}, HAT-P-57 \citep{2015AJ....150..197H} or KOI-976 \citep{2019AJ....158...88A}, represent a small fraction of the entire exoplanet population \citep[see e.g.][]{2021AJ....162..204H}. On the one hand, these systems are analogous to those binary stars, where the pulsation of one of the stars is influenced by the tidal forces of its companion, such as V453 Cyg \citep{2020MNRAS.497L..19S} or RS Cha \citep{2021A&A...645A.119S}. Planet-to-star interactions of this kind have been observed in several of these systems, including HAT-P-2 \citep{2017ApJ...836L..17D}, WASP-33 \citep{2022A&A...660L...2K} and HD 31221 \citep{2023arXiv230504000K}. Recently, \cite{2024arXiv240308014B} explored the potential of these resonances in revealing the evolution of exoplanets, primarily the highly eccentric HAT-P-2b. The emerging class of interacting pulsators and their planetary companions is expected to yield new opportunities in stellar characterisation as well \citep{2024arXiv240308014B}. On the other hand, due to the strong irradiation, these systems are prime candidates for atmospheric studies via phase curve analyses. The ultra-hot Jupiter (UHJ) WASP-33b, being the epitome of these systems, was the subject of many such studies, including analyses based on the occultation detections \citep[e.g.][]{2011MNRAS.416.2096S,2012ApJ...754..106D,2015A&A...584A..75V,2015ApJ...806..146H} and the full phase curve \citep[e.g.][]{2018AJ....155...83Z}. Both of these require careful treatment of the stellar pulsations, which is enabled by the continuous photometry done by spaceborne observatories such as \texttt{Kepler} \citep{2010Sci...327..977B} or Transiting Exoplanet Survey Satellite \citep[\texttt{TESS;}][]{2015JATIS...1a4003R}, as seen in \cite{2020A&A...639A..34V}. \cite{2021A&A...645A.119S} explored the known $\delta$ Scuti population in the \texttt{Kepler} field and found no UHJ companions of these stars. This suggests that these objects are rare and all-sky surveys, such as \texttt{TESS} or \texttt{PLATO} \citep[PLAnetary Transits and Oscillations of stars][]{2014ExA....38..249R} might provide a more in-depth insight into these objects.

The host stars in question typically present either $\delta$ Scuti, $\gamma$ Doradus, or hybrid $\delta$ Scuti -- $\gamma$ Doradus pulsations. These objects often rotate rapidly \citep{2022ApJ...928...35A}, leading to spin-orbit misalignment detectable by Doppler tomography -- a technique also pioneered on the WASP-33 system \citep{2010MNRAS.407..507C, 2015ApJ...810L..23J}. A continuous monitoring of the stellar obliquity $\lambda$ also allows for the detection of nodal precession \citep[e.g.][]{2015ApJ...810L..23J, 2020MNRAS.492L..17S, 2022ApJ...931..111S, 2022MNRAS.512.4404W}.

\begin{figure*}[!h]
     \centering
     \includegraphics[width = \textwidth]{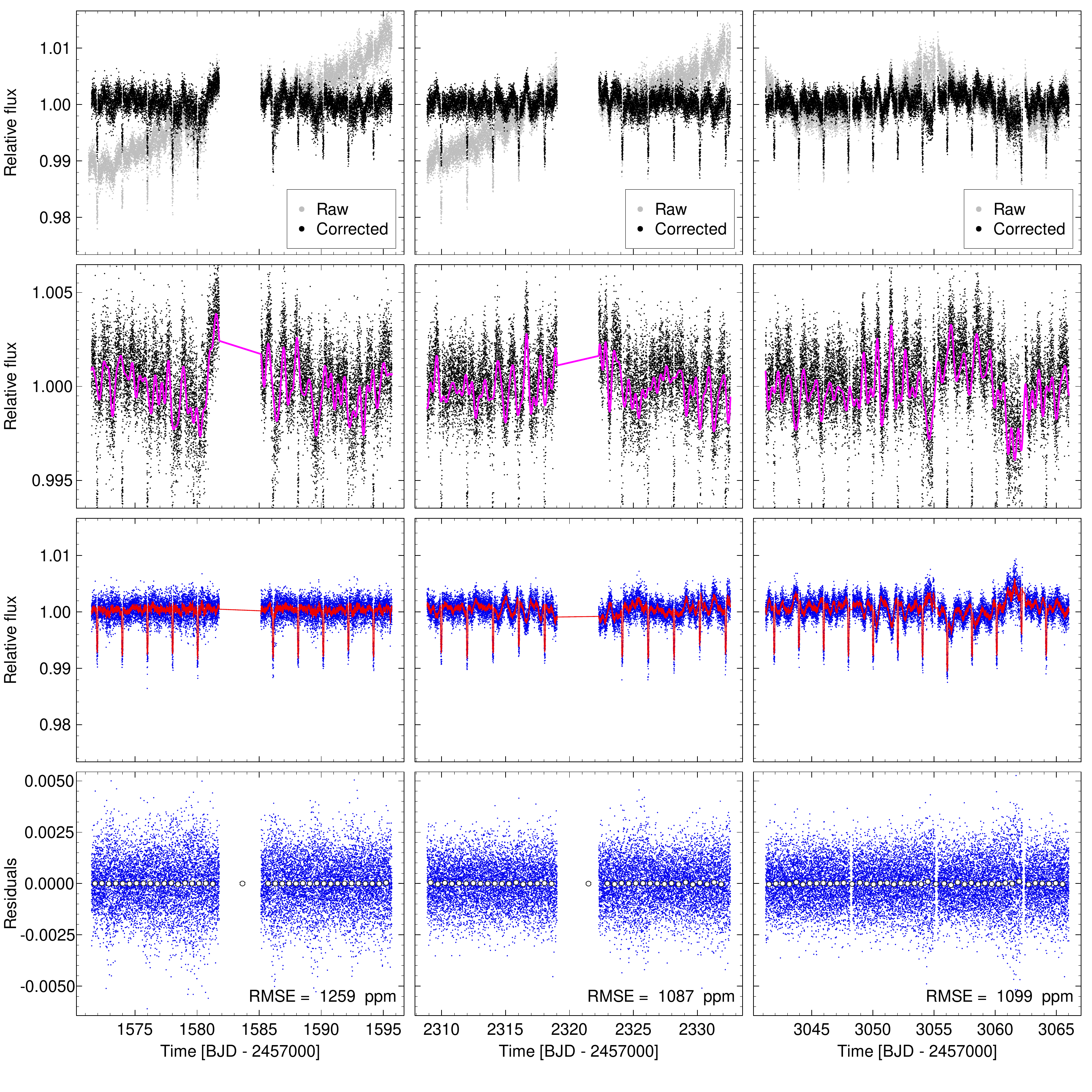}
     \caption{TESS Simple Aperture Photometry light curve of WASP-167 (top row, grey dots) from Sectors 10 (left), 37 (middle), and 64 (right). The masked, detrended and centroid-corrected LC is shown with black. The pulsation model, constructed from the 16 highest amplitude sinusoidal curves, is shown in the second row with a solid magenta line. Blue dots in the third row represent the LC when these 16 highest amplitude pulsations have been removed, while the best-fit combination of time-correlated noise and phase curve models are shown with red. The residuals, along with their Root Mean Square Error (RMSE) are shown in the bottom row.}
     \label{fig:inputlc}
 \end{figure*}

In the WASP-167 system (also known as KELT-13, TOI-748,\break TIC 104024556, Gaia DR3 6154982877300947840), \cite{2017MNRAS.471.2743T} discovered a hot Jupiter with an orbital period of $2.0219596 \pm 0.0000006$ days and a radius of $1.58 \pm 0.05$~R$_J$. The mass measurement of WASP-167b through radial velocities (RVs) was inconclusive, as the star is both `line-poor' and `broad-lined', owing to its spectral type ($T_{\rm eff} = 6900 \pm 150$~K) and rapid rotation \citep[$v \sin i_\star = 52 \pm 8$~km~s$^{-1}$;][]{2017MNRAS.471.2743T}. Based on Doppler tomography, \cite{2017MNRAS.471.2743T} established that WASP-167b/KELT-13b has a misaligned, retrograde orbit, determined by the spin-orbit angle $\lambda = -165 \pm 5^\circ$. The line-profile tomography also led the authors to conclude that the host star presents pulsations of either $\delta$ Scuti or $\gamma$ Doradus type, which presents further difficulties in the mass measurements to the ones discussed above \citep[see e.g.][]{2015A&A...578L...4L}.

We undertake an analysis of WASP-167b, based solely on its \texttt{TESS} light curve (LC). This paper is structured as follows. In Sect.~\ref{sec:methods}, we describe the LC pre-processing and fitting. In Sect.~\ref{sec:results}, we present the best-fit parameters, describing both the transits and the out-of-transit variations, as well as give constraints on the mass of WASP-167b. We also explore how the pulsations of the host affect the determination of these parameters. We characterise the pulsations of the host star, discuss the possibilities of planet-to-star interactions and the possible effects of the pulsations on the precision and accuracy of the retrieved parameters in Sect.~\ref{sec:discussion}. We also estimate the dayside and nightside temperatures of WASP-167b in Sect. \ref{sec:discussion}. 

 \section{Methods}\label{sec:methods}
\subsection{Light curve preparation} \label{sec:lcprep}

We downloaded the short-cadence Simple Aperture Photometry (SAP) light curves obtained in Sectors 10, 37 and 64 by \texttt{TESS} using the \texttt{lightkurve} software \citep{2018ascl.soft12013L}, utilising the \texttt{astroquery} and \texttt{astropy} packages \citep{2019AJ....157...98G, 2022ApJ...935..167A}. We removed all data points with a non-zero quality flag, and corrected for the third light contamination via the \texttt{CRWDSAP} keyword. The obtained light curves are shown in Fig. \ref{fig:inputlc}. We then proceeded to mask apparent anomalies and remove a linear trend from the data of all three sectors. We also corrected for the jitter in the pointing of the telescope. This was done by subtracting the best-fit linear model of the centroid position of the PSF (Point Spread Function) of WASP-167 in terms of row and column pixels.

\begin{figure*}
    \centering
    \includegraphics[width = \textwidth]{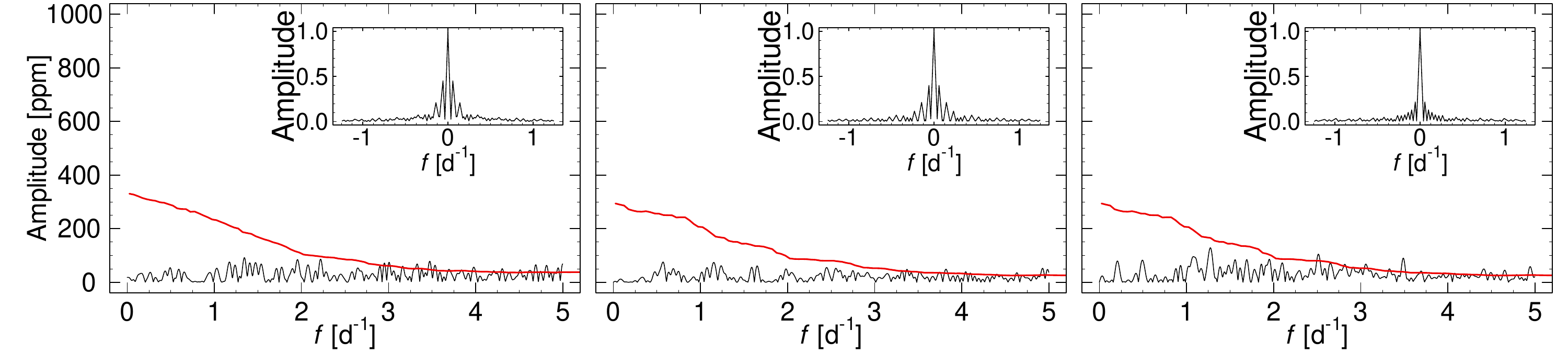}
    \caption{Fourier spectrum after subtracting the 16 highest-amplitude sinusoidal curves for Sectors 10 (left), 37 (middle), and 64 (left). The solid red line corresponds to the noise level as estimated by \texttt{Period04}. The window function are shown in the top right corner of each panel.}
    \label{fig:residuals_fourier}
\end{figure*}

After the initial preparatory steps, we examined the Fourier spectra of the light curves using \texttt{Period04} \citep{2005CoAst.146...53L}. We identified a `critical frequency' ($\sim 0.98$ d$^{-1}$, labelled F2 in Table \ref{tab:freqs}, see Sect \ref{sec:discussion} for more details) that is close to being in resonance with the second orbital harmonic of WASP-167b. To explore the effect of the stellar pulsations  on the stability of the phase curve parameters, 
we prepared 17 different light curves, 16 of which have been pre-whitened by the 16 highest-amplitude sinusoidal signals from the LCs of the three sectors and the original one (i.e. without whitening). To do this,  we first masked the transits. We then computed the Fourier spectrum of the LCs. \cite{1993A&A...271..482B} suggest that in order to consider a peak in the power spectrum significant, an S/N $\geq$ 4 is required. Relying on this criterion is common practise in the analysis of pulsating stars \citep[e.g.][]{2009MNRAS.394..995D, 2017MNRAS.464.1553D, 2018ApJS..237...15A, 2020A&A...640A..36B, 2022A&A...667A..60S, 2023ApJS..269...32S}. We identified $16$ peaks with S/N $\geq$ 4 using \texttt{Period04}, separately for all three sectors. The whitened light curves are then given as
\begin{equation} \label{eq:prep}
    \Phi_i(t) = \Phi (t)  - \sum_{i = 0}^{16} A_i \sin \left(2\pi \left(\nu_i \cdot t + \varphi_i \right) \right), 
\end{equation}
where $\Phi (t)$ is the original detrended SAP light curve, and $A_i$, $\nu_i$ and $\varphi_i$ are the amplitudes, frequencies and phases of the individual Fourier-terms and their values are taken from the Fourier-analysis. We emphasise that the pre-whitening process is done for each Sector individually. The light curve corresponding to $\Phi_{16}$ is shown in Fig. \ref{fig:inputlc}, and it contains $32$ transits and $32$ full occultations as we fitted the data from the three sectors jointly. We note that the sole purpose of this initial Fourier analysis is to enable a consistent phase curve modelling. The Fourier spectrum after the subtraction of the first $16$ highest-amplitude sinusoidal curves is shown on Fig. \ref{fig:residuals_fourier}. For the more in-depth discussion of the astrophysical aspect of the pulsations (Sect. \ref{sec:discussion}), we need a stable LC model that can take advantage of a dataset that does not need masked transits.

\subsection{Analysis of the light curves}
\begin{figure*}
    \centering
    \includegraphics[width = \textwidth]{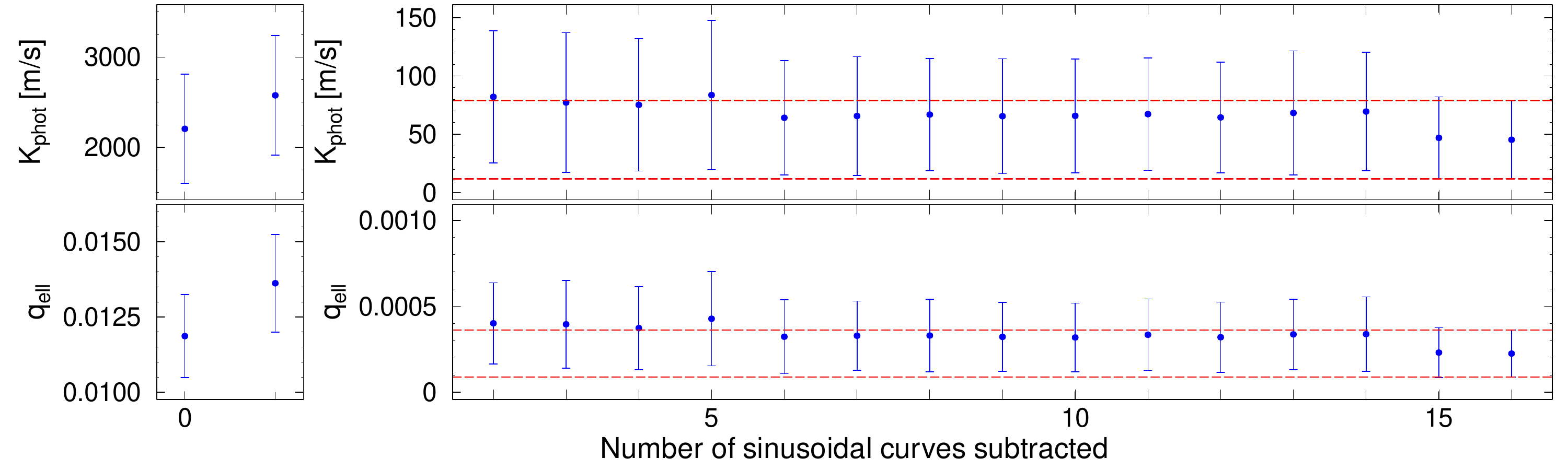}
    \caption{Best-fit $K_{\rm phot}$ and $q_{\rm ell}$ as a function of the number of sinusoidal curves subtracted from the light curve. Note the change in the scale of the y axis after subtracting one sinusoidal curve (i.e. before and after the break). The dashed lines correspond to the uncertainty of the parameters obtained from the $\Phi_{16}(t)$ LC.}
    \label{fig:Kq}
\end{figure*}
We made use of the Transit and Light Curve Modeller \citep[\texttt{TLCM};][]{2020MNRAS.496.4442C} to model the light variations induced by the presence of WASP-167b. \texttt{TLCM} uses a model similar to the one of \cite{2002ApJ...580L.171M} to describe the transits. It is characterised by the relative planetary radius ($R_{\rm p}/R_\star$); the impact parameter which is related to the orbital inclination via $b = a/R_\star \cos i_{\rm p}$ ($i_{\rm p}$ being the orbital inclination of the planet relative to the line of sight); the time of mid-transit ($t_C$) and the scaled semi-major axis ($a/R_\star$). The stellar parameters listed in Table \ref{tab:stellarparams} were taken into account via the isochrones of \cite{2000MNRAS.315..543H}. We calculated the theoretical values for the $u_1$ and $u_2$ quadratic limb-darkening coefficients in the \texttt{TESS} passband with \texttt{PyLDTK} \citep{Parviainen2015}, which uses the \texttt{PHOENIX} library of stellar atmospheres \citep{Husser2013}. The stellar parameters used in these calculations were adopted from \citet{2017MNRAS.471.2743T} and are listed in Table \ref{tab:stellarparams}. We implemented a reparametrization of quadratic limb-darkening law (based on Principal Component Analysis), by fitting the two coefficients $A$ and $B$, defined as
\begin{align} 
    A &= \frac{1}{4}\left( u_+ \left( \frac{1}{\alpha} - \frac{1}{\beta} \right) + u_- \left( \frac{1}{\alpha} + \frac{1}{\beta} \right) \right), \label{eq:quadldA}\\
    B &= \frac{1}{4}\left( u_+ \left( \frac{1}{\alpha} + \frac{1}{\beta} \right) + u_- \left( \frac{1}{\alpha} - \frac{1}{\beta} \right) \right) \label{eq:quadldB},
\end{align}
where $u_{\pm} = u_1 \pm u_2$, $\alpha = \frac{1}{2} \cos \left( 77^\circ \right)$ and $\beta = \frac{1}{2} \sin \left( 77^\circ \right)$. Note that {\sc TLCM} uses this definition of limb darkening coefficients in its latest version. We also applied Gaussian priors on $A$ and $B$, around their theoretical values, in the form of $\mathcal{N}(0.91, 0.10)$ and $\mathcal{N}(1.40, 0.10)$.

\begin{table}
\caption{Stellar parameters of WASP-167, adopted from \cite{2017MNRAS.471.2743T}.}
\label{tab:stellarparams}
\centering
\scriptsize
\begin{tabular}{c c c c c }
\hline
\hline
$T_{\rm eff}$ [K] & $\log g$ & [Fe/H] & $v \sin I_\star$ [km s$^{-1}$] & $R_\star$ [$R_\odot$]  \\
$6900 \pm 150$ & $4.13 \pm 0.02$ & $0.1 \pm 0.1$ & $52 \pm 8$  & $1.79 \pm 0.05$   \\
\hline
\end{tabular}
\end{table}
We modelled the out-of-transit variations including the Doppler beaming, the ellipsoidal variability \citep[][and references therein]{2007ApJ...670.1326Z, 2011MNRAS.415.3921F}, and the reflection effect, as described in \cite{2020MNRAS.496.4442C} and \cite{2021arXiv210811822C}. Both the Doppler beaming and the ellipsoidal variations can be used to estimate the mass of WASP-167b, as the former is characterised by the (photometric) RV semi-amplitude $K_{\rm phot}$, while the latter introduces the ellipsoidal mass ratio, $q_{\rm ell}$, as a fitting parameter. The contribution of these two effects to the photometric baseline variations depend on the stellar parameters and are taken into account according to the description in \cite{2020MNRAS.496.4442C}. The reflection of the stellar light from the planetary atmosphere was taken into account via a Lambertian phase function:
\begin{equation}\label{eq:phc}
    \frac{F_{\rm p}}{F_\star} = \frac{I_{\rm p}}{I_\star} \left( \frac{R_{\rm p}}{R_\star} \right)^2 + A_g  \left( \frac{R_{\rm p}}{R_\star} \frac{R_\star}{a} \right)^2 \frac{\sin \alpha - \alpha \cos \alpha}{\pi},
\end{equation}
where $I_{\rm p}/I_\star$ is the surface brightness ratio of the planet and the star, $A_g$ is the geometric albedo, and $F_{\rm p}$ and $F_\star$ are the planetary and stellar fluxes, respectively. The phase angle $\alpha$ is related to the true anomaly $v$ by
\begin{equation}
    \cos \left( \alpha + \varepsilon \right) = \cos \left( \omega + v \right) \sin i_{\rm p},
\end{equation}
where $\omega$ is the argument of periastron and $\varepsilon$ is the offset of the brightest point on the planetary surface from the substellar point. We assumed that WASP-167b has a circular orbit, as expected for a hot Jupiter. Additional free parameter was the orbital period $P$. Furthermore, we applied the so-called height correction, denoted by the fitting parameter $h$ \citep[see Eq. (47) of][]{2020MNRAS.496.4442C}. This parameter describes an offset in flux that is designed to perform renormalisation in a similar way to third light contamination.  

\subsection{Time-correlated noise}

In their systematic phase curve study of the \texttt{TESS} data, \cite{2020AJ....160..155W} excluded WASP-167 specifically due to the stellar variability, which introduces difficulties to the analysis of both the transits and the out-of-transit variations. One possible way to account for this was presented in the case of WASP-33 by \cite{2020A&A...639A..34V}, where the $\delta$ Scuti/$\gamma$ Doradus type stellar oscillations had been removed from the light curve as simple sinusoidal signals, computed via a periodogram. \texttt{TLCM} incorporates the wavelet-based routines of \cite{2009ApJ...704...51C} to account for time-correlated noise \citep{2020MNRAS.496.4442C, 2021arXiv210811822C}. By treating the residual stellar oscillations and the remaining systematic effects (Fig. \ref{fig:inputlc}) in this way, we are able to improve the precision and accuracy of the fitted parameters, as was demonstrated in \cite{2021arXiv210811822C,2022arXiv220801716K, 2023A&A...674A.186B}. Stellar pulsations in the case of WASP-33 were also accounted for successfully via this noise filtering approach in \cite{2022A&A...660L...2K}. These routines introduce two additional parameters to describe the noise: $\sigma_r$ and $\sigma_w$ for the red and white components, respectively. 

We used these wavelets to account for any systematic effects and residual stellar pulsations for the 16 pre-whitened light curves and the original one. In every case, we ran the MCMC analysis for a maximal number of 100,000 steps, with 10 chains (also known as walkers) and a thinning factor of 10.

\section{Results}\label{sec:results}

\subsection{Influence of the stellar pulsations on the out-of-transit variability}

\begin{figure*}
    \centering
    \includegraphics[width = \textwidth]{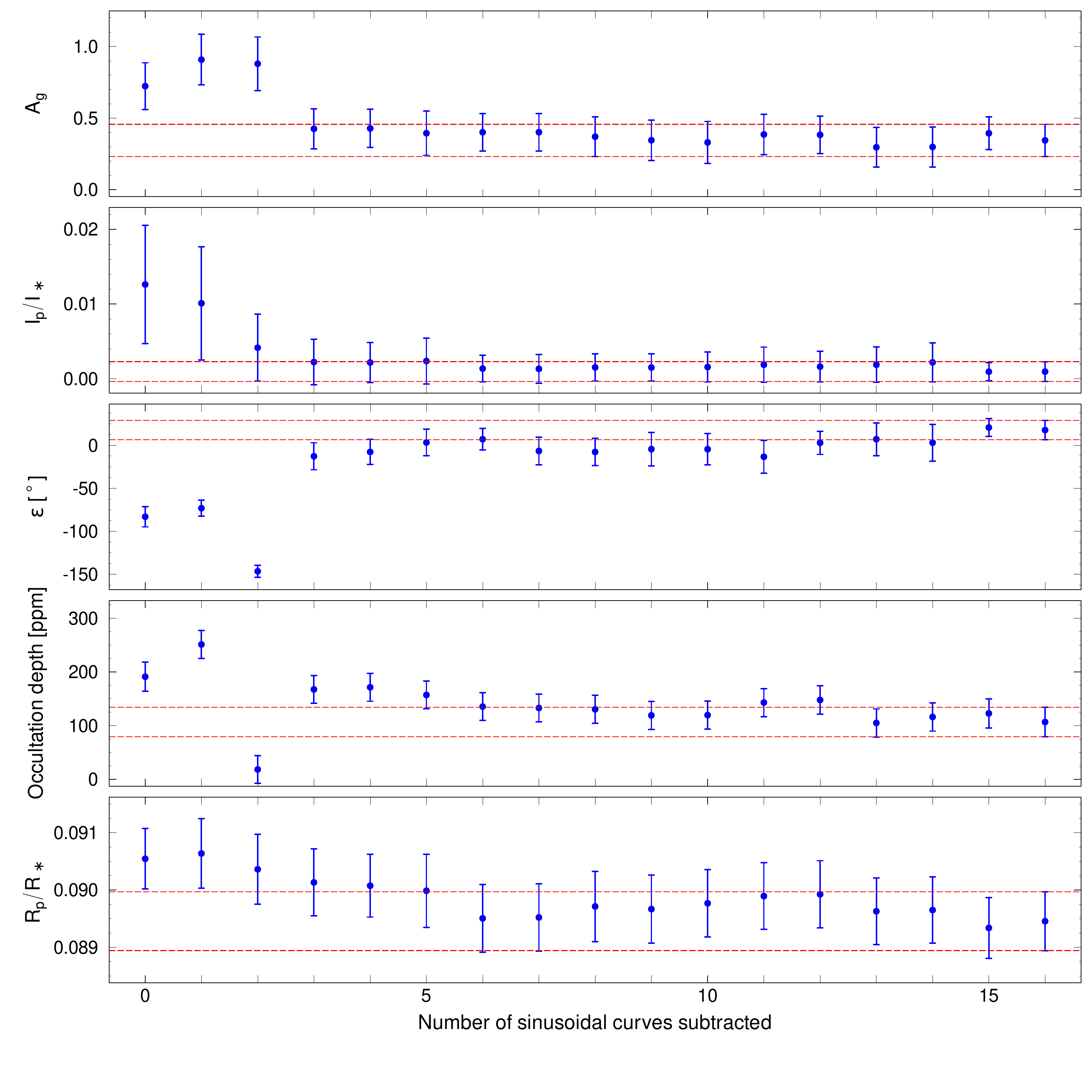}
    \caption{Best-fit  $A_g$, $I_{\rm p}/I_\star$, $\varepsilon$, occultation depth and $R_{\rm p}/R_\star$ as a function of the number of sinusoidal curves subtracted from the light curve. The dashed lines correspond to the $1 \sigma$ uncertainty of the parameters obtained from the $\Phi_{16}(t)$ LC.}
    \label{fig:consts}
\end{figure*}


In order to explore the effect that the stellar pulsation has on the precision and accuracy of the out-of-transit variations, we show the behaviour of $A_g$, $I_{\rm p}/I_\star$, $\varepsilon$, the occultation depth, $K_{\rm phot}$, and $q_{\rm ell}$ as a function of the number of sinusoidal curves removed as described in  Eq. (\ref{eq:prep}). As a check for consistency, we also show the estimated $R_{\rm p}/R_\star$ in every case. The parameters are shown in Figs. \ref{fig:consts}. The occultation depth can be calculated as follows. First, we  divide the best-fit model (including the transits, occultations and the out-of-transit variations) with the sum of Eq. (\ref{eq:phc}), the ellipsoidal distortion and the Doppler boosting. The median flux level during the (total) occultation can then be compared to the median of the flat baseline that is out-of-transit and out-of-occultation. The uncertainty of the occultation depth can be estimated as the standard deviation divided by the square root of the number of LC points during the occultation and the standard deviation divided by the square root of the number of points out-of-transit and out-of-occultation, added in quadrature.

In \texttt{TLCM}, the Doppler beaming and ellipsoidal variations are fitted simultaneously to constrain each other \cite{2020MNRAS.496.4442C}. As discussed in Sect. \ref{sec:lcprep}, one component of the stellar variability is close to second orbital harmonic of WASP-167b -- the same time-dependence as the ellipsoidal effect. The wavelet-based noise filtering of \cite{2009ApJ...704...51C}, as implemented in \texttt{TLCM} \cite{2020MNRAS.496.4442C, 2021arXiv210811822C}, `force' the LC into the most likely shape. Given the additional signal around the $2/P$ frequency region, this component of the stellar variability can yield an inaccurate (either underestimated or overestimated) mass for WASP-167b. Theoretically, it would be possible to rely on the Doppler boosting alone for the mass estimation \citep{2012A&A...541A..56M}. However, as \cite{2020MNRAS.496.4442C} pointed out, that effect is often degenerate with the model for the light reflected from the planetary atmosphere. Although \cite{2020MNRAS.496.4442C} suggests the usage of radial velocity data to break this degeneracy between the out-of-transit parameters, the available measurements from \cite{2017MNRAS.471.2743T} imply that for this system, the mass of the companion can not be determined with adequate precision from the present data. The two parameters representing these two  effects (beaming and ellipsoidal variations), $K_{\rm phot}$ and $q_{\rm ell}$ do not change significantly after subtracting the sinusoidal curve corresponding to the frequency near $2/P$, labelled F2 in Table \ref{tab:freqs} (Fig. \ref{fig:Kq}, Table \ref{tab:oot_params}), however, they are consistent with each other when modelling $\Phi_2(t)$ -- $\Phi_{16}(t)$.

When we model $\Phi_2(t)$, the offset between the brightest point on the stellar surface and the substellar point ($\varepsilon$) is $<-90^\circ$ i.e. on the nightside of the planet, assuming tidal locking. This case also corresponds to the most shallow occultation (Fig.~\ref{fig:consts}, Table~\ref{tab:oot_params}).  This value of $\varepsilon$ is nonphysical. All of the best-fit parameters are consistent with each other when extracted from $\Phi_3(t)$ -- $\Phi_{16}(t)$ (Figs.~\ref{fig:Kq} and \ref{fig:consts}). The marginal detection of an eastward offset\footnote{According to \cite{2021arXiv210811822C}, positive $\varepsilon$ values correspond to eastward shift, while negative values imply a westward shift, under the assumption that the planet has a prograde rotation.} between the brightest point on the planetary surface and the substellar point observed in $\Phi_{15}(t)$ and $\Phi_{16}(t)$ are within $1.8\sigma$ to the respective values from $\Phi_3(t)$ and within $2\sigma$ to $0^\circ$.

To decide which set of parameters to accept, we calculated the Bayesian Information Criteria (BIC) for all 17 light curve solutions (Table \ref{tab:oot_params}). When calculating the BIC values, we used the best-fit model without the correlated noise identified via the wavelets in comparison to $\Phi_{0}(t)$ -- $\Phi_{16}(t)$. The total number of parameters was therefore $13 + 3i$, where $i$ is the number of sinusoidal curves subtracted -- that are described by three independent parameters: their frequency, amplitude and phase -- (without $\sigma_w$ and $\sigma_r$).  The BIC analysis suggests that the $\Phi_{16}(t)$ light curve produces the optimal set of parameters, we therefore adopt this one. We note that $R_{\rm p}/R_\star$ does not change significantly across the $17$ tested cases. This is expected, since the transits have higher S/N than the out-of-transit variability, and the transit depths can therefore be estimated with higher precision and accuracy in comparison to other parameters \citep{2021arXiv210811822C}. We further emphasise that the uncertainties of the analysed parameters are also consistent with each other (Table \ref{tab:oot_params}). 

The uniform priors (also known as searchboxes) used in the analysis for each individual parameter were the same throughout the analysis of the 17 light curves. These are shown in Table~\ref{tab:sysparams}. The combination of the best-fit model for the planetary transits, occultations, out-of-transit variability and the red noise (for $\Phi_{16}(t)$) is plotted in Fig.~\ref{fig:inputlc}.

\subsection{System parameters} \label{sec:syspars}

\begin{table*}
\caption{Fitted and derived parameters from the $\Phi_{16}(t)$ light curve.}
\label{tab:sysparams}
\centering
\begin{tabular}{l c c c c}
\hline
\hline
Parameter & Searchbox & This work & This work (zero mass) & \cite{2017MNRAS.471.2743T}\\
\hline
$a / R_\star$ &  [$0.1$, $9.9$] & $4.126 \pm 0.054$ & $4.122 \pm  0.053$
 & $4.38 \pm 0.36$ \\ 
 $R_{\rm p} / R_\star$ & [$0.0$, $0.2$] & $0.08945 \pm 0.00051$ & $0.08953 \pm 0.00053$& $0.0906 \pm 0.0043$ \\ 
 $b$ & [$0.0$, $1.0$] & $0.7627 \pm 0.0089$ &$0.7635 \pm 0.0086$
 &$0.77 \pm 0.01$ \\ 
 $P$ [days] &  [$2.017$, $2.027$]& $2.02195830 \pm 0.00000053$ & $2.02195831   \pm        0.00000056$
&$2.0219596 \pm 0.0000006$ \\ 
 $t_C$ [BTJD] & [$8592.13$, $8592.23$] & $8592.18253 \pm 0.00027$ & $8592.18254 \pm  0.00028$
 &$6592.4643 \pm 0.0002$\\ 
$A$\tablefootmark{a} & [$-1.5$, $1.5$] & $0.86 \pm 0.10$ & $0.85 \pm 0.10$
 & -- \\ 
$B$\tablefootmark{b} & [$-1.5$, $1.5$] & $1.291 \pm 0.083$ & $1.290 \pm 0.082$
& --\\ 
 $\sigma_{r}$ [100 ppm] & [$0.0$, $10000.0$] & $703.1 \pm 7.1$ &$703.1 \pm 7.2$ & -- \\ 
 $\sigma_{w}$ [100 ppm] & [$0.0$, $6000.0$] & $10.527 \pm 0.034$ & $10.526 \pm 0.034$  & -- \\ 
$A_g$ & [$0.0$, $1.0$] & $0.34 \pm 0.11$ & $0.32 \pm 0.12$ &-- \\ 
$I_p/I_\star$ & [$0.0$, $1.0$] & $0.0010 \pm 0.0013$ & $0.0025 \pm 0.0027$ &--\\ 
$\varepsilon$  [$^\circ$]& [$-90.0$, $90.0$] & $17.7 \pm 11.2$ &  $18.2 \pm 12.0$ & --\\ 
$K_{\rm phot}$  [m s$^{-1}$]& [$0.0$, $200000.0$] & $45 \pm 34$ & --& --\\ 
$K$  [m s$^{-1}$]& -- & -- & --&$< 897$\tablefootmark{c} \\ 
$q_{\rm ell}$ & [$0.0$, $0.4$] & $0.00022 \pm 0.00014$ & --&--\\ 
$h$ & [$-0.5$, $0.5$] & $-0.00055 \pm 0.00015$ & $-0.00055 \pm 0.00015$ &-- \\
\hline
\multicolumn{4}{c}{Derived parameters} \\
\hline
\multicolumn{2}{l}{$R_\star$ [$R_\odot$]} & $1.861 \pm 0.057$  & $1.84 \pm 0.17$ & $1.79 \pm 0.05$ \\
\multicolumn{2}{l}{$M_\star$ [$R_\odot$]} & $1.49 \pm 0.13$  & $1.44 \pm  0.32$ &$1.59 \pm 0.08$\\
\multicolumn{2}{l}{$R_{\rm p}$ [$R_{\rm J}$]} & $1.621 \pm  0.059$ & $1.60 \pm 0.16$& $1.58 \pm 0.05$\\
\multicolumn{2}{l}{$i_p$ [$^\circ$]} & $79.3 \pm 0.2$ & $79.3 \pm 0.2$ & $79.9 \pm 0.3$ \\
\multicolumn{2}{l}{$u_1$} & $0.453 \pm 0.062$ &  $0.453 \pm 0.062$ & -- \\
\multicolumn{2}{l}{$u_2$} & $0.030 \pm 0.067$ &  $ 0.029 \pm 0.069$ & -- \\
\multicolumn{2}{l}{$M_{\rm P, Beaming}$ [$M_{\rm J}$]} & $0.37 \pm 0.28$& -- & $<8$\tablefootmark{c}\\
\multicolumn{2}{l}{$M_{\rm P, Ellipsoidal}$ [$M_{\rm J}$]} & $0.34 \pm 0.22$ &-- &  --\\
\multicolumn{2}{l}{Occultation depth [ppm]} & $106.8 \pm 27.3$ & $108.8 \pm 27.3$ &$<1100$ ppm\\
\hline
\end{tabular}
\tablefoot{\tablefootmark{a}{Gaussian prior applied: $\mathcal{N}(0.91,0.10)$.} \tablefootmark{b}{Gaussian prior applied: $\mathcal{N}(1.40,0.10)$.}\tablefootmark{c}{Estimated from radial velocities, a method similar in principle to Doppler beaming.}}
\end{table*}

\begin{figure}
    \centering
    \includegraphics[width = \columnwidth]{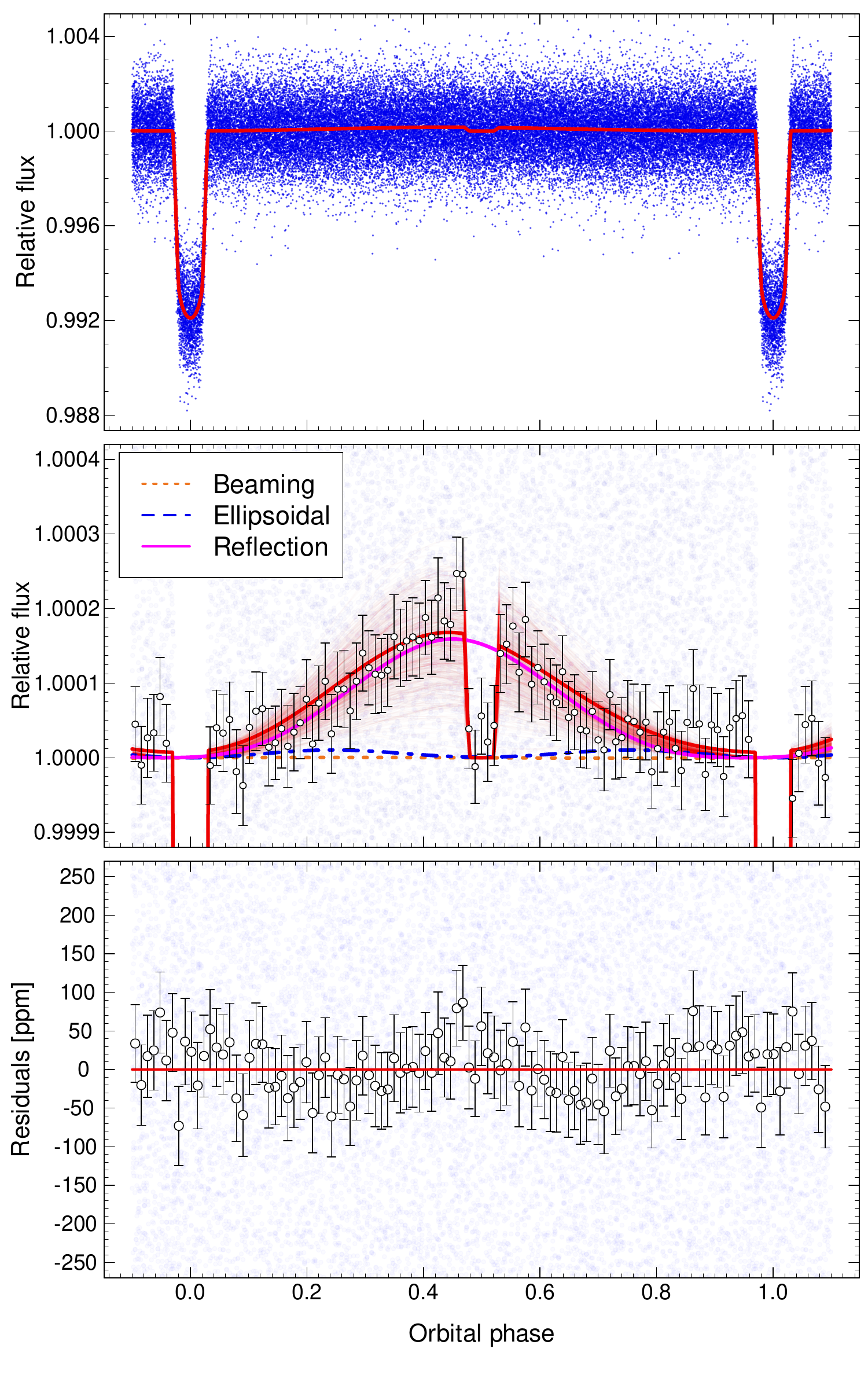}
    \caption{Full phase curve model of WASP-167b (solid red line). The top panel shows the $\Phi_{16}(t)$ light curve with the fitted red noise subtracted via the wavelet-based filtering algorithm. The occultation is clearly visible on the zoomed-in middle panel, with the reflection effect, ellipsoidal variability and Doppler beaming separated (magenta, blue, and orange lines, respectively). The residuals are shown on the bottom panel. White circles and the respective error bars represent bins of $500$ data points each. The less opaque red lines correspond to $1000$ model LCs are randomly selected from the ($1\sigma$) uncertainty ranges of each of the parameters.}
    \label{fig:occ}
\end{figure}

We applied the same uniform priors for the free parameters (Table \ref{tab:sysparams}) on all 17 light curves, with the exception of the two limb darkening coefficients, where the same Gaussian priors were used. The best-fitting light curve model, obtained from $\Phi_{16}(t)$ is shown in Fig. \ref{fig:occ} and the corresponding parameters are shown in Table \ref{tab:sysparams}. The basic transit parameters, including $a/R_\star$, $R_{\rm p}/R_\star$ and $b$, agree within $1\sigma$ with the respective values presented in the discovery paper of \citep{2017MNRAS.471.2743T} -- after converting those to the formalism used in our analysis. A slight ($\lesssim 1.7 \sigma$) discrepancy can be observed in the measurements of $i_{\rm p}$, which may be attributed to the different treatment of limb darkening. The results of \cite{2017MNRAS.471.2743T} are based on the four parameter law \citep{2000A&A...363.1081C, 2004A&A...428.1001C} with theoretical coefficients, the applicability of which is disputable for moderately fast rotating, pulsating stars, such as WASP-167. It is also possible that the transit chord is moving due to nodal precession, as in the cases of WASP-33b\break \citep[e.g.][]{2015ApJ...810L..23J, 2022ApJ...931..111S}, KELT-9b \citep{2022ApJ...931..111S} or Kepler-13Ab \citep[e.g.][]{2014MNRAS.437.1045S, 2018AJ....155...13H, 2020MNRAS.492L..17S}, The planetary radius of $1.621 \pm 0.059$~R$_J$ is also in good agreement with the size reported in \cite{2017MNRAS.471.2743T}. We also present slight improvement in the precision of these parameters from the \texttt{TESS} data alone. 

The well-known degeneracies between the scaled semi-major axis, the relative planetary radius and the impact parameter are observable in Fig. \ref{fig:corner}. These can be quantified via the correlation coefficients from the posteriors: $r = -0.49$ (between $a/R_\star$ and $R_{\rm p}/R_\star$), $-0.95$ (between $a/R_\star$ and $b$), and $0.51$ (between $R_{\rm p}/R_\star$ and $b$). The reparametrization of the quadratic limb-darkening coefficients, expressed in Eqs. (\ref{eq:quadldA}) and (\ref{eq:quadldB}) yield no significant degeneracies between these and the transit parameters discussed above. 

\cite{2017MNRAS.471.2743T} provide an upper limit of $8$ M$_J$ on the mass of WASP-167 by analysing the radial velocity curve. Our adopted set of solutions suggest that the mass is considerably lower at $0.37 \pm 0.28$~M$_J$ or $0.34 \pm 0.22$~M$_J$, as obtained by modelling the Doppler beaming effect and the ellipsoidal variability, respectively, placing the planet in the mass regime of Saturn. We note however that due to the stellar signal near the second orbital harmonic (Fig. \ref{fig:fou}), these masses, obtained from the marginal detection of the two effects, should not be taken at face value. Furthermore, the correlation between the posteriors (Fig. \ref{fig:corner}) of $K_{\rm phot}$ and $q_{\rm ell}$ ($r = 0.83$) also implies that caution must be taken when considering the mass of this object. The true mass of WASP-167b may be estimated by modeling these two effects from infrared light curve observations, where the stellar activity is known to be less significant. 

The observed occultation depth of $106.8 \pm 27.3$ ppm is well below the upper limit of $1100$ ppm in $z'$ band, as obtained from \texttt{TRAPPIST} observations \citep{2017MNRAS.471.2743T} \citep[see][for comparison of the passbands]{2015JATIS...1a4003R}. Incidentally, this result also emphasises the necessity of space-based observations in the study of exoplanets. This occultation depth is detected with $p = 1.12 \cdot 10^{-6}$, based on the two-sample t-test in which the out-of-transit and out-of-occultation flat baseline is compared to the signal level during the occultation. We also measure a slight eastward offset in the brightest point of the planetary surface at $17.7^\circ \pm 11.2^\circ$, a value that is similar to the observed by \cite{2020A&A...639A..34V} for WASP-33b. The posteriors of the parameters describing the observed light variation of WASP-167b are not correlated with each other, or the mass measuring parameters (Fig.~\ref{fig:corner}).

The wavelet-based noise fitting in \texttt{TLCM} is constrained by prescribing that the standard deviation of the residuals (of the LC and red noise models) must be equal to the average photometric uncertainty \citep{2021arXiv210811822C}. This implies that the efficiency of the noise treatment is dependent on, among other things, the brightness of the star (in this case, $V = 10.52$ mag). Accordingly, some autocorrelated effects may be present in the residuals, as seen in Fig. \ref{fig:occ}. To rule out the possibility that the residuals seen in Fig. \ref{fig:occ} are caused by an overestimation of the ellipsoidal variability (and consequently, the mass of WASP-167b),\break we ran another test with the beaming and ellipsoidal effects turned off (i.e. a mass-less case) on the $\Phi_{16}$ LC. The resultant parameters are also shown in Table \ref{tab:sysparams}. All of the best-fit parameters are in excellent agreement with the case when we also modelled the two mass-related effects. We measure an increase in $I_p/I_\star$, which is a consequence of the lack of a (near) constant signal that would have been the sum of the beaming and ellipsoidal effects. The surface brightness ratio is therefore enlarged (yet still consistent with the uncertainty range from the original analysis) to account for the underlying occultation depth (which is observed to be $108.8 \pm 27.3$~ppm). The resultant (`zero mass') LC model is shown in Fig. \ref{fig:occ_nomass}. The same auto-correlated effects are present in the residuals of this modelling as in the case of Fig. \ref{fig:occ}, implying that the mass of WASP-167b is not overestimated in the original calculations. Consequently, we adopt the more thorough `massive' case for the rest of the analysis.

As an additional check for consistency, we also modelled the light curves from Sectors 10, 37 and 64 separately, for all 17 stages of pre-whitening. We fitted the same parameters, over the same intervals as in the global analysis (Table \ref{tab:sysparams}). The resultant parameters (Table \ref{tab:params_comp_sep_sec}) are shown on Figs. \ref{fig:kq_sep_sec} and \ref{fig:consts_sep_sec}. None of the parameters extracted in this way differ by more than $3\sigma$ from the adopted ones. The mass-related parameters (Fig. \ref{fig:kq_sep_sec}) from Sector 10 are always consistent with a zero-mass object. The effect of the `critical frequency' F2 is, however, present in the $K_{\rm phot}$ and $q_{\rm ell}$ values from Sectors 37 and 64. As in the case of the global analysis, the relevant values from $\Phi_2$ to $\Phi_{16}$ are also consistent with zero in all three Sectors. Certain $\varepsilon$ values are non-physical (\ref{fig:consts_sep_sec}), yielding negative occultation depths (as in the case of $\Phi_{15}$ and $\Phi_{16}$ from Sector 37). These are, however, consistent with $0$ within the estimated uncertainty range (Table \ref{tab:params_comp_sep_sec}), implying that the out-of-transit variability can not be distinguished from the time-correlated noise in these cases.  The pre-whitening process is not perfectly efficient at the removal of all quasi-periodic noise features (Fig. \ref{fig:inputlc}). Consequently, its residuals (third row of \ref{fig:inputlc}) show variability on timescales comparable to the orbital period, but much shorter than the duration of the light curve of the three combined sectors. As a result, in our global fit, these quasi-periodic events get `averaged out', yielding parameters that are fundamentally more precise and accurate. This phenomenon can also be observed in the case of WASP-33b, where \cite{2015A&A...578L...4L} used a large number of radial velocity observations to deal with the effects of stellar pulsations (in contrast with \cite{2013A&A...553A..44K}, who only had access to $12$ data points in total). We note that because of the deep transits (yielding high S/N for the transit depth), the  $R_{\rm p}/R_\star$ are always consistent with each other within $1 \sigma$. We also note that in general, Sector 10 yield the highest uncertainties for any given parameter, corresponding to the highest RMSE seen in the data from these three Sectors (Fig. \ref{fig:inputlc}).

We further note that the height correction parameter $h$ is not degenerated with any other fitting parameter (Fig. \ref{fig:corner}). This implies that its use is effective in treating any possible light contamination that is not corrected for via the \texttt{CROWDSAP} keyword during the photometry.

\section{Discussion}\label{sec:discussion}
\subsection{Stellar signal}

\begin{table}[!h]
\caption{Frequencies, amplitudes, phases and the respective S/N values extracted from the Fourier transform.}
\label{tab:freqs}
\centering
\scriptsize 
\begin{tabular}{l c c c c}
\hline
\hline
 & Frequency [$d^{-1}$] & Amplitude & Phase & S/N\\
\hline
\multicolumn{5}{c}{Sector 10} \\
\hline
F1 & $0.0831 \pm 0.0005$ & $810.0 \pm 17.2$ & $0.5519 \pm 0.0034$ & $44.6$ \\ 
F2 & $0.9874 \pm 0.0006$ & $650.0 \pm 17.2$ & $0.2938 \pm 0.0042$ & $27.8$ \\ 
F3 & $0.6235 \pm 0.0009$ & $435.9 \pm 17.2$ & $0.6381 \pm 0.0063$ & $20.7$ \\ 
F4 & $0.3009 \pm 0.0006$ & $618.5 \pm 17.2$ & $0.3291 \pm 0.0044$ & $30.4$ \\ 
F5 & $0.8305 \pm 0.0008$ & $493.5 \pm 17.2$ & $0.3118 \pm 0.0055$ & $22.7$ \\ 
F6 & $1.7426 \pm 0.0013$ & $307.6 \pm 17.2$ & $0.9985 \pm 0.0089$ & $11.7$ \\ 
F7 & $0.3591 \pm 0.0008$ & $503.4 \pm 17.2$ & $0.0406 \pm 0.0054$ & $25.1$ \\ 
F8 & $0.8781 \pm 0.0008$ & $476.3 \pm 17.2$ & $0.7507 \pm 0.0057$ & $21.9$ \\ 
F9 & $0.7350 \pm 0.0011$ & $358.2 \pm 17.2$ & $0.5646 \pm 0.0076$ & $16.8$ \\ 
F10 & $1.6400 \pm 0.0012$ & $317.0 \pm 17.2$ & $0.6849 \pm 0.0086$ & $12.4$ \\ 
F11 & $2.1232 \pm 0.0017$ & $227.8 \pm 17.2$ & $0.6201 \pm 0.0120$ & $7.6$ \\ 
F12 & $0.5293 \pm 0.0014$ & $274.5 \pm 17.2$ & $0.4919 \pm 0.0100$ & $12.9$ \\ 
F13 & $1.2725 \pm 0.0017$ & $228.2 \pm 17.2$ & $0.2254 \pm 0.0120$ & $8.7$ \\ 
F14 & $0.2112 \pm 0.0018$ & $216.4 \pm 17.2$ & $0.3051 \pm 0.0126$ & $11.1$ \\ 
F15 & $1.8348 \pm 0.0019$ & $205.8 \pm 17.2$ & $0.4257 \pm 0.0133$ & $7.5$ \\ 
F16 & $1.5909 \pm 0.0019$ & $210.5 \pm 17.2$ & $0.2256 \pm 0.0130$ & $8.2$ \\ 
F17 & $1.0819 \pm 0.0031$ & $124.8 \pm 17.2$ & $0.6277 \pm 0.0219$ & $5.1$ \\ 
F18 & $2.0317 \pm 0.0032$ & $120.5 \pm 17.2$ & $0.3283 \pm 0.0227$ & $4.1$ \\ 
F19 & $2.5274 \pm 0.0034$ & $113.9 \pm 17.2$ & $0.9292 \pm 0.0240$ & $3.7$ \\ 
F20 & $0.7661 \pm 0.0020$ & $191.9 \pm 17.2$ & $0.5987 \pm 0.0143$ & $8.9$ \\ 
F21 & $0.0083 \pm 0.0010$ & $376.4 \pm 17.2$ & $0.7295 \pm 0.0073$ & $22.1$ \\ 
F22 & $1.3414 \pm 0.0033$ & $118.9 \pm 17.2$ & $0.2652 \pm 0.0230$ & $4.5$ \\ 
\hline
\multicolumn{5}{c}{Sector 37} \\
\hline
F1 & $0.8537 \pm 0.0001$ & $438.4 \pm 14.5$ & $0.2611 \pm 0.0005$ & $244.8$ \\ 
F2 & $0.9818 \pm 0.0005$ & $664.9 \pm 14.5$ & $0.8651 \pm 0.0035$ & $37.0$ \\ 
F3 & $0.2145 \pm 0.0008$ & $438.8 \pm 14.5$ & $0.8865 \pm 0.0053$ & $25.0$ \\ 
F4 & $0.1748 \pm 0.0009$ & $360.4 \pm 14.5$ & $0.2597 \pm 0.0064$ & $20.8$ \\ 
F5 & $1.7489 \pm 0.0006$ & $571.8 \pm 14.5$ & $0.8042 \pm 0.0040$ & $26.0$ \\ 
F6 & $1.8180 \pm 0.0007$ & $512.4 \pm 14.5$ & $0.2650 \pm 0.0045$ & $23.1$ \\ 
F7 & $0.4225 \pm 0.0012$ & $274.5 \pm 14.5$ & $0.5237 \pm 0.0084$ & $14.8$ \\ 
F8 & $0.0463 \pm 0.0014$ & $235.9 \pm 14.5$ & $0.8155 \pm 0.0098$ & $17.4$ \\ 
F9 & $2.1126 \pm 0.0016$ & $215.3 \pm 14.5$ & $0.9823 \pm 0.0107$ & $9.6$ \\ 
F10 & $1.0427 \pm 0.0013$ & $251.6 \pm 14.5$ & $0.6853 \pm 0.0092$ & $13.4$ \\ 
F11 & $1.6955 \pm 0.0011$ & $308.3 \pm 14.5$ & $0.7244 \pm 0.0075$ & $14.0$ \\ 
F12 & $1.4170 \pm 0.0025$ & $134.2 \pm 14.5$ & $0.4598 \pm 0.0172$ & $6.5$ \\ 
F13 & $0.7817 \pm 0.0017$ & $197.1 \pm 14.5$ & $0.5396 \pm 0.0117$ & $10.6$ \\ 
F14 & $0.2704 \pm 0.0017$ & $202.0 \pm 14.5$ & $0.1712 \pm 0.0114$ & $10.7$ \\ 
F15 & $0.3733 \pm 0.0025$ & $135.3 \pm 14.5$ & $0.4879 \pm 0.0170$ & $7.2$ \\ 
F16 & $2.2255 \pm 0.0032$ & $104.5 \pm 14.5$ & $0.7059 \pm 0.0221$ & $4.9$ \\ 
F17 & $3.0125 \pm 0.0034$ & $98.9 \pm 14.5$ & $0.3743 \pm 0.0233$ & $4.4$ \\ 
F18 & $0.8548 \pm 0.0001$ & $381.8 \pm 14.5$ & $0.2452 \pm 0.0006$ & $213.2$ \\ 
F19 & $1.8874 \pm 0.0025$ & $135.7 \pm 14.5$ & $0.1838 \pm 0.0170$ & $6.0$ \\ 
F20 & $1.5438 \pm 0.0035$ & $96.1 \pm 14.5$ & $0.6740 \pm 0.0240$ & $4.5$ \\ 
F21 & $3.4828 \pm 0.0042$ & $80.8 \pm 14.5$ & $0.2763 \pm 0.0285$ & $3.8$ \\ 
F22 & $0.5737 \pm 0.0040$ & $84.6 \pm 14.5$ & $0.2580 \pm 0.0273$ & $4.7$ \\ 
\hline
\multicolumn{5}{c}{Sector 64} \\
\hline
F1 & $0.1114 \pm 0.0004$ & $843.0 \pm 13.9$ & $0.7294 \pm 0.0026$ & $34.1$ \\ 
F2 & $0.9874 \pm 0.0005$ & $654.4 \pm 13.9$ & $0.2529 \pm 0.0034$ & $20.7$ \\ 
F3 & $0.8437 \pm 0.0004$ & $691.2 \pm 13.9$ & $0.6012 \pm 0.0032$ & $23.2$ \\ 
F4 & $0.1589 \pm 0.0005$ & $584.4 \pm 13.9$ & $0.4518 \pm 0.0038$ & $22.3$ \\ 
F5 & $0.0690 \pm 0.0004$ & $732.1 \pm 13.9$ & $0.4386 \pm 0.0030$ & $33.7$ \\ 
F6 & $0.3003 \pm 0.0006$ & $513.9 \pm 13.9$ & $0.4419 \pm 0.0043$ & $19.5$ \\ 
F7 & $0.4490 \pm 0.0008$ & $406.3 \pm 13.9$ & $0.0158 \pm 0.0054$ & $15.4$ \\ 
F8 & $1.8261 \pm 0.0008$ & $390.5 \pm 13.9$ & $0.0408 \pm 0.0057$ & $9.8$ \\ 
F9 & $1.7409 \pm 0.0009$ & $325.6 \pm 13.9$ & $0.8936 \pm 0.0068$ & $8.5$ \\ 
F10 & $0.2676 \pm 0.0007$ & $434.2 \pm 13.9$ & $0.3894 \pm 0.0051$ & $16.2$ \\ 
F11 & $0.9031 \pm 0.0009$ & $344.6 \pm 13.9$ & $0.4086 \pm 0.0064$ & $11.3$ \\ 
F12 & $0.7871 \pm 0.0010$ & $317.7 \pm 13.9$ & $0.5489 \pm 0.0069$ & $10.6$ \\ 
F13 & $1.6481 \pm 0.0012$ & $250.0 \pm 13.9$ & $0.0324 \pm 0.0088$ & $6.8$ \\ 
F14 & $0.5599 \pm 0.0012$ & $264.0 \pm 13.9$ & $0.6380 \pm 0.0084$ & $9.3$ \\ 
F15 & $0.3661 \pm 0.0015$ & $207.4 \pm 13.9$ & $0.4260 \pm 0.0106$ & $8.1$ \\ 
F16 & $0.7305 \pm 0.0014$ & $225.3 \pm 13.9$ & $0.3250 \pm 0.0098$ & $7.5$ \\ 
F17 & $1.3476 \pm 0.0017$ & $185.0 \pm 13.9$ & $0.3958 \pm 0.0119$ & $5.5$ \\ 
F18 & $2.1325 \pm 0.0016$ & $197.5 \pm 13.9$ & $0.1294 \pm 0.0112$ & $5.0$ \\ 
F19 & $2.2308 \pm 0.0022$ & $141.8 \pm 13.9$ & $0.2955 \pm 0.0156$ & $3.7$ \\ 
F20 & $0.6110 \pm 0.0017$ & $180.7 \pm 13.9$ & $0.4660 \pm 0.0122$ & $6.3$ \\ 
F21 & $0.6792 \pm 0.0019$ & $160.2 \pm 13.9$ & $0.1076 \pm 0.0138$ & $5.5$ \\ 
F22 & $1.2744 \pm 0.0021$ & $144.0 \pm 13.9$ & $0.5849 \pm 0.0153$ & $4.4$ \\ 
\hline
\end{tabular}
\end{table}

\begin{figure*}
    \centering
    \includegraphics[width = \textwidth]{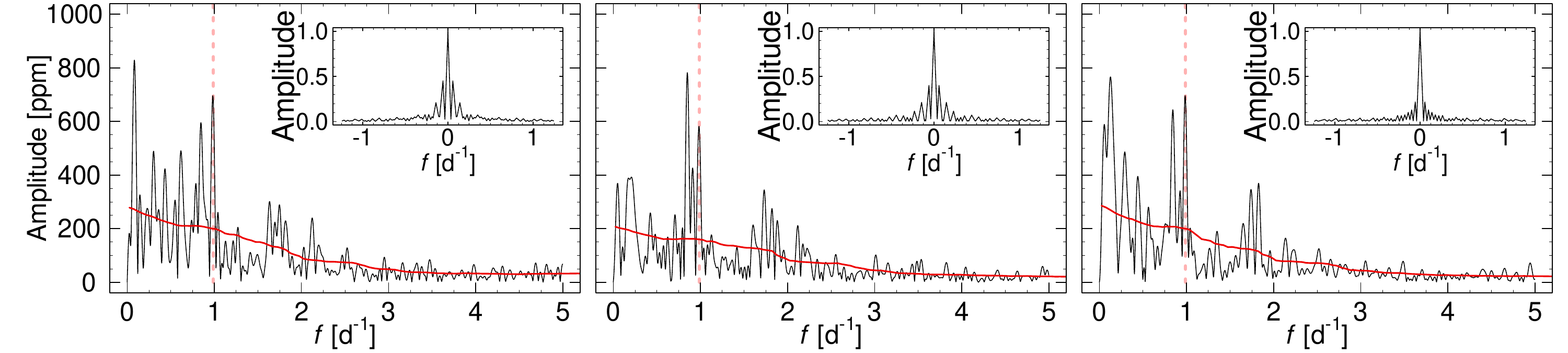}
    \caption{Fourier spectrum of the residuals of the LC model of WASP-167 from Sectors 10, 37, and 64 (left, middle and right panel). The inset plots show the spectral windows. The dashed red line marks the second orbital harmonic of WASP-167b, demonstrating the `critical frequency' that influences primarily the mass measurements of the planet. The noise level, as calculated by \texttt{Period04}, is shown with a solid red line.}
    \label{fig:fou}
\end{figure*}

The analysis of the stellar pulsations generally requires uninterrupted light curves. To this end, we subtracted the phase curve model (including the transits, occultations and out-of-transit effects) obtained by the analysis of $\Phi_{16}(t)$ from $\Phi_0(t)$. We then proceeded with calculating the Fourier spectrum of these residuals using \texttt{Period04}, separately for the LCs of the three Sectors. These spectra are shown in Fig. \ref{fig:fou}. The frequencies, amplitudes and phases, that are the result of the Fourier analysis, are presented in Table \ref{tab:freqs}, for peaks with S/N $\gtrsim 4$. In all three Sectors, we extracted data from $22$ peaks with the highest amplitudes. While a detailed analysis of the stellar signal is beyond the scope of this paper, several observations can be made from Fourier spectra. ($i$) There are no significant peaks in the data beyond $\sim 3.5~{\rm d^{-1}}$. ($ii$) There is a peak in the Fourier spectra -- F2 from Table \ref{tab:freqs} -- that almost exactly coincides with the second orbital harmonic of WASP-167b in all three Sectors (Fig. \ref{fig:fou}). When comparing to the best-fit orbital frequency ($f_{\rm orb} = (0.49547000 \pm 1.3 \cdot 10^{-7})$~d$^{-1}$), we find that F2 can be expressed as $1.996$, $1.958$, and $1.996$ times $f_{\rm orb}$ in the three Sectors, respectively. Similar commensurability was noted by \cite{2014A&A...561A..48V} and later confirmed by \cite{2022A&A...660L...2K} in the case of the WASP-33 system. ($iii$) The frequencies and amplitudes of the spectral peaks vary considerably from Sector to Sector, with the exception of F2, the `critical frequency'. Given the incompatible nature of the Fourier spectra in the three Sectors, searching for  a phase shift in the individual sinusoidal curves\citep[e.g.][]{2014A&A...561A..48V} is not feasible.

We find that $2f_{\rm orb}$ matches F2 within $\approx 3 \sigma$ (in Sectors 10 and 64), and within $15 \sigma$ in Sector 37, when using the formal uncertainties of the pulsational frequencies, calculated via \texttt{Period04}. These calculations are based on the idealistic case where the pulsations can be expressed as simple sinusoidal signals with constant phases, amplitudes and frequencies. As we demonstrate in this paper, this is not the case for WASP-167, and, as a consequence, these uncertainties may be considerably underestimated. We constructed a simple test to investigate the possibility that the `critical frequency' only appears close to $2f_{\rm orb}$ without underlying tidal interactions. To that end, similarly to \cite{2014A&A...561A..48V} and \cite{2023arXiv230504000K}, we constructed $10^5$ synthetic `frequency spectra', each consisting of 22 frequencies (corresponding to the number of peaks identified and presented in Table \ref{tab:freqs}), drawn from a uniform distribution between $0$~d$^{-1}$ and $3.5$~d$^{-1}$. We then proceeded to match these to $2f_{\rm orb}$ to within $3 \cdot 0.0015$~d$^{-1}$, or roughly the average uncertainty in the frequencies from Table \ref{tab:freqs}, thus providing a conservative estimate for the synthetic F2. We found that commensurability defined in this way only arises in $5.6\%$ of the cases. This statistic further serves as evidence for tidal interactions between WASP-167b and its host.


The frequency range in which the stellar activity presents itself means that the observing windows might not be sufficient for a detailed characterisation of its nature. If it originates from pulsations, as stated in \cite{2017MNRAS.471.2743T}, then WASP-167b is a $\gamma$ Doradus pulsator. Indeed, instability for this type of stellar pulsation may be expected on the time scales available through the \texttt{TESS} photometry \citep[e.g.][]{2014A&A...568A..34B,2015MNRAS.450.3015K}.  Although they are generally not expected for $T_{\rm eff} \approx 7000$ K, stellar spots can also produce variability compatible with our observations (Fig. \ref{fig:fou}) and they might offer a better explanation for the apparent changes in the frequencies and amplitudes that are seen in the Fourier decomposition. It is also possible that the star presents both $\gamma$ Dor type pulsations and spots. Future studies focusing on the Zeeman effect \citep{1947ApJ...105..105B}, or more precise photometric observations with longer baselines  \citep[with e.g. \texttt{PLATO}][]{2014ExA....38..249R} may help understanding the origins of the stellar variability of WASP-167. We note, however, that the accurate interpretation of the stellar signal depends strongly on the phase curve model of the planet (Fig. \ref{fig:occ}), as the latter is also comprised of sinusoidal terms with comparable amplitudes to the ones from stellar oscillational frequencies (Table \ref{tab:freqs}).

The peak in the Fourier spectra near the second orbital harmonic (Fig. \ref{fig:fou}) is likely composed of both stellar variability (regardless of its nature) and the ellipsoidal effect. As discussed in Sect. \ref{sec:syspars}, this implies that the mass measurements of WASP-167b\break might be inaccurate. This issue might be solved by obtaining more, higher quality radial velocity data than that presented in \cite{2017MNRAS.471.2743T}, as in the case of WASP-33 \citep{2015A&A...578L...4L}. It is also possible that, as stated in Sect. \ref{sec:syspars}, IR observations by e.g. \texttt{Ariel} \citep{2021arXiv210404824T} may be able to provide more reliable mass constraints, since the pulsational amplitudes are known to be lower at higher wavelengths \citep[e.g.][]{2009MNRAS.394..995D}.

\begin{figure*}
    \centering
    \includegraphics[width = \textwidth]{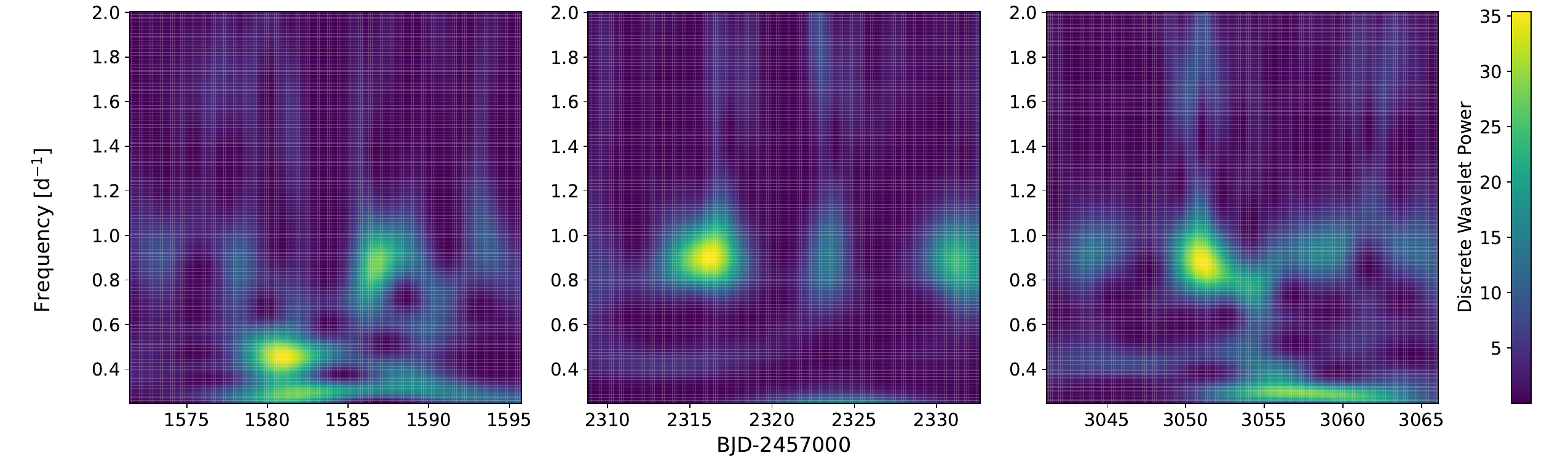}
    \caption{Wavelet maps of the stellar activity from Sectors 10 (left), 37 (middle) and 64 (right).}
    \label{fig:vw}
\end{figure*}

We also see that the F2 frequency is present in the Fourier spectra from all three Sectors (Fig. \ref{fig:fou}), with similar amplitudes. This suggest that WASP-167b is influencing the activity of its host. To examine the intra-sector stability, we used a time-frequency method known as the discrete wavelet transform \citep{1996AJ....112.1709F} after filling in the data downlink gaps (seen in Fig. \ref{fig:inputlc}) with zeros. The wavelet maps of Fig. \ref{fig:vw} show the $0.25$ -- $2$ d$^{-1}$ frequency regime (thus incorporating F2). There is clear amplitude modulation around F2, with a period of $\sim 4$ days, or $\sim 2P$. A similar behaviour has also been observed in the case of WASP-33b\break and it's host star \cite{2022A&A...660L...2K}. This timescale is well above the estimated orbital period $\leq 1.81$~d$^{-1}$, suggesting that the origin of this particular signal is not related to stellar rotation. We also observe that in the $\leq 0.8$~d$^{-1}$ frequency regime there are considerable amplitude and phase curve modulations both intra-sector, and from Sector to Sector. However, the long-term effects, such as detector systematics, would also be observable in this range, and they can not be distinguished from the stellar activity from the light curves used in this analysis.  A deeper exploration of the stellar signal would need to carefully remove all of the known instrumental effects from the light curves, however, such a study is beyond the scope of this paper. Instrumental noise sources could therefore be a possible explanation for the inter-Sector variability in the Fourier spectrum (Fig. \ref{fig:fou} and Table~\ref{tab:freqs}). Another possibility might be the instability of the pulsational or spot patterns of the star, at least on the time scales at which the three observed Sectors sample the LC.


\subsection{Dayside, nightside and intrinsic temperature of WASP-167b}

In the phase curve parametrization of Eq. (\ref{eq:phc}), the second term, characterised by the geometric albedo, is made up of light that is reflected from the planetary atmosphere (determined by the Bond albedo, $A_B$) as well as light that is emitted from the dayside. To constrain the dayside and nightside temperatures, we use the toy model presented in \cite{kelt9pow} and used in \cite{companion} for the energy budget of WASP-167b, to distinguish between the two phenomena. In a thermodynamical equilibrium, we can state that the incoming and outgoing luminosities are related to each other by:
\begin{equation}
    L_{\rm income} + L_{\rm intrinsic} = L_{\rm day} + L_{\rm night} + L_{\rm absorbed}.
\end{equation}
In this formulation, $L_{\rm intrinsic}$ is the result of the intrinsic planetary heat. The incident luminosity is related to the bolometric stellar luminosity via
\begin{equation}
    L_{\rm income} = L_\star \left( \frac{R_{p}}{2a} \right)^2.
\end{equation}
By definition, the reflected light from the planetary surface can be characterised as
\begin{equation}
    L_{\rm reflected} = (1-A_B) L_{\rm absorbed}.
\end{equation}
Assuming blackbody radiation for the dayside and nightside hemispheres of the planet, we get:
\begin{align}
    T_{\rm day} &= \sqrt[4]{T_{\rm intrinsic}^4 + \frac{1}{2}f(1-A_B) T_{\rm eff}^4 \left( \frac{R_\star}{a} \right)^2}, \\
    T_{\rm night} &= \sqrt[4]{T_{\rm intrinsic}^4 + \frac{1}{2}(1-f)(1-A_B) T_{\rm eff}^4 \left( \frac{R_\star}{a} \right)^2}.
\end{align}
Here $T_{\rm intrinsic}$, $T_{\rm day}$ and $T_{\rm night}$ are the internal, dayside and nightside temperature of WASP-167b, respectively. The heat redistribution efficiency, $f$, determines the amount of incident stellar luminosity that passes through to the nightside of the planet.

The amplitude of the phase curve in Eq. (\ref{eq:phc}) is the combination of emitted and reflected light. By neglecting the marginal eastward offset, the theoretical geometric albedo can be characterised as
\begin{equation}
    A_{\rm g,~theo}(A_{\rm B}, f, T_{\rm int})  = \frac{1}{2}\left(\frac{a}{R_\star}\right)^2 \frac{I_{\rm day}  -I_{\rm night}}{I_\star}+\frac{1}{4} A_B \left(\frac{R_\star}{a}\right)^2,
\end{equation}
where $I_{\rm day}$ are the passband-specific luminosities of a unit-sized surface area. The surface brightness ($I_{\rm p}/I_\star$) ratio of Eq.~(\ref{eq:phc}), taken from $\Phi_{16}(t)$ (Table \ref{tab:sysparams}) can be matched to $I_{\rm night}/I_\star$.

There are therefore two observable quantities and three parameters ($A_B$, $f$ and $T_{\rm intrinsic}$) that need to be determined from them. Thus, no unique solution can be found for these via observations in a single passband. We therefore carried out a Monte Carlo integration by generating a sample of $2\cdot 10^6$ independent random values for $A_B$, $f$ and $T_{\rm intrinsic}$, within intervals of $[0,1]$, $[0.5,1]$ and $[0\ \text{K},5000\ \text{K}]$, respectively, with a uniform distribution. The uncertainties of $T_{\rm eff}$ and $a/R_\star$ (Tables \ref{tab:stellarparams} and \ref{tab:sysparams}) were taken into account via an additional $2 \cdot 10^6$ independent random values, with Gaussian distributions (with means and standard deviations taken from Tables \ref{tab:stellarparams} and \ref{tab:sysparams}). We then proceeded by matching $A_{g, \text{theo}}$ and $I_{\rm night}/I_\star$ to the $2\sigma$ uncertainty ranges for $A_g$ and $I_{\rm p} / I_\star$ from Table \ref{tab:sysparams}.

\begin{figure*}
    \centering
    \includegraphics[width = \textwidth]{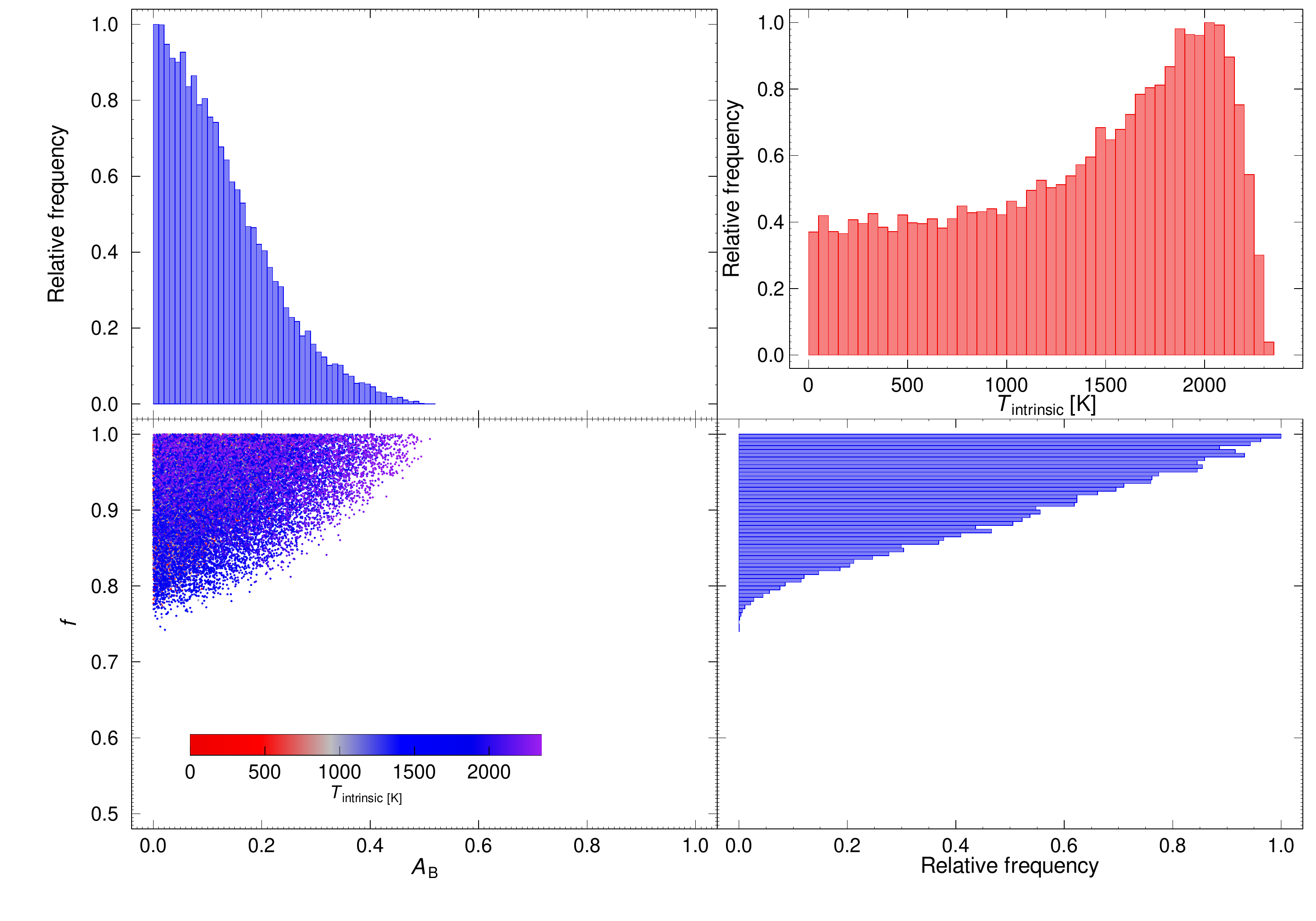}
    \caption{Distribution of the Bond albedo, the heat redistribution parameter and the intrinsic temperature, the combination of which reproduces the observed phase curve of WASP-167b.}
    \label{fig:AfT}
\end{figure*}

\begin{figure}
    \centering
    \includegraphics[width = \columnwidth]{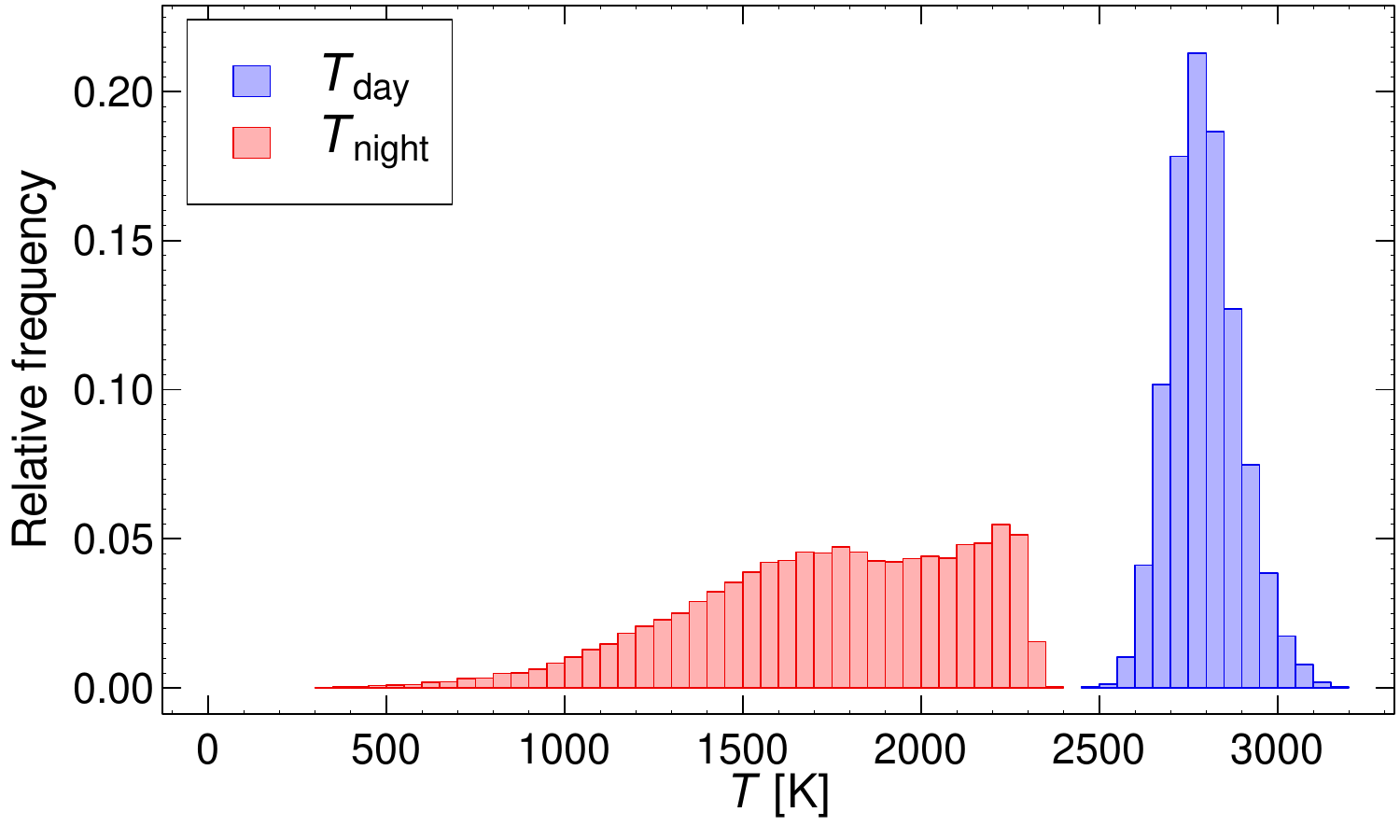}
    \caption{Distribution of the dayside and nightside temperatures that reproduce the observed phase curve of WASP-167b.}
    \label{fig:daynight}
\end{figure}

The resultant distributions of the Bond albedo, intrinsic temperature and the heat redistribution efficiency are shown in Fig. \ref{fig:AfT}. From these, we can draw the following conclusions. \mbox{WASP-167b/KELT-13b} has a moderately low Bond albedo, with $A_B < 0.51$ at the $2\sigma$ level. This is generally in line with what is expected from a hot Jupiter based on empirical studies \citep[e.g.][]{2008ApJ...689.1345R, 2009ApJ...703..769K,  2011ApJ...729...54C, 2020NatAs...4..453D, 2023arXiv231103264S, companion} and theoretical predictions \citep[e.g.][]{2000ApJ...538..885S}. We also place a lower limit on the heat redistribution efficiency of $f > 0.74$ (at $2\sigma$), implying that the atmosphere of the planet is structured in a way that it retains the majority of the incident (and absorbed) stellar flux on the dayside, similarly to WASP-18Ab \citep{2011ApJ...742...35N}. We find that the intrinsic temperature of WASP-167b is $\leq 2340$~K, with $1470 \pm 640$~K representing the median and standard deviation of the distribution shown in Fig.~\ref{fig:AfT}. Combining these facts with the $5.6$~L$_\odot$ stellar luminosity (in the \texttt{TESS} passband) implies that the dayside temperature of the planet (Fig. \ref{fig:daynight}) is $2790 \pm 100$~K, implying that WASP-167b is a so-called ultra hot Jupiter. We also place an upper limit on the nightside temperature (Fig. \ref{fig:daynight}) of $2360$~K, with the uncertainty range estimated as $1780 \pm 380$~K. In fact, given that our $I_{\rm p}/I_\star$ estimates are consistent with $0$ in the majority of the cases (Fig \ref{fig:consts}), it is safe to assume that all of the atmospheric parameters provided here can be regarded as upper limits. We note that $A_B$, $f$ and $T_{\rm intrinsic}$ are heavily degenerate, which is the consequence of analysing data in a single passband. Similarly to the nature of the stellar signal, future observations with e.g. \verb|PLATO| or \verb|Ariel| will have the possibility of a more in-depth exploration of the atmospheric temperature regimes of this planet.

We emphasise that all of the solutions represented on Fig. \ref{fig:AfT} satisfy the requirements for $A_g$ and $I_p/I_\star$ equally well, and that no solutions were found beyond the limits for $A_B$ and $f$ presented above. The $2 \sigma$ uncertainty range stems from taking the $2 \sigma$ uncertainties for $A_g$ and $I_p/I_\star$. 

\begin{figure}
    \centering
    \includegraphics[width = \columnwidth]{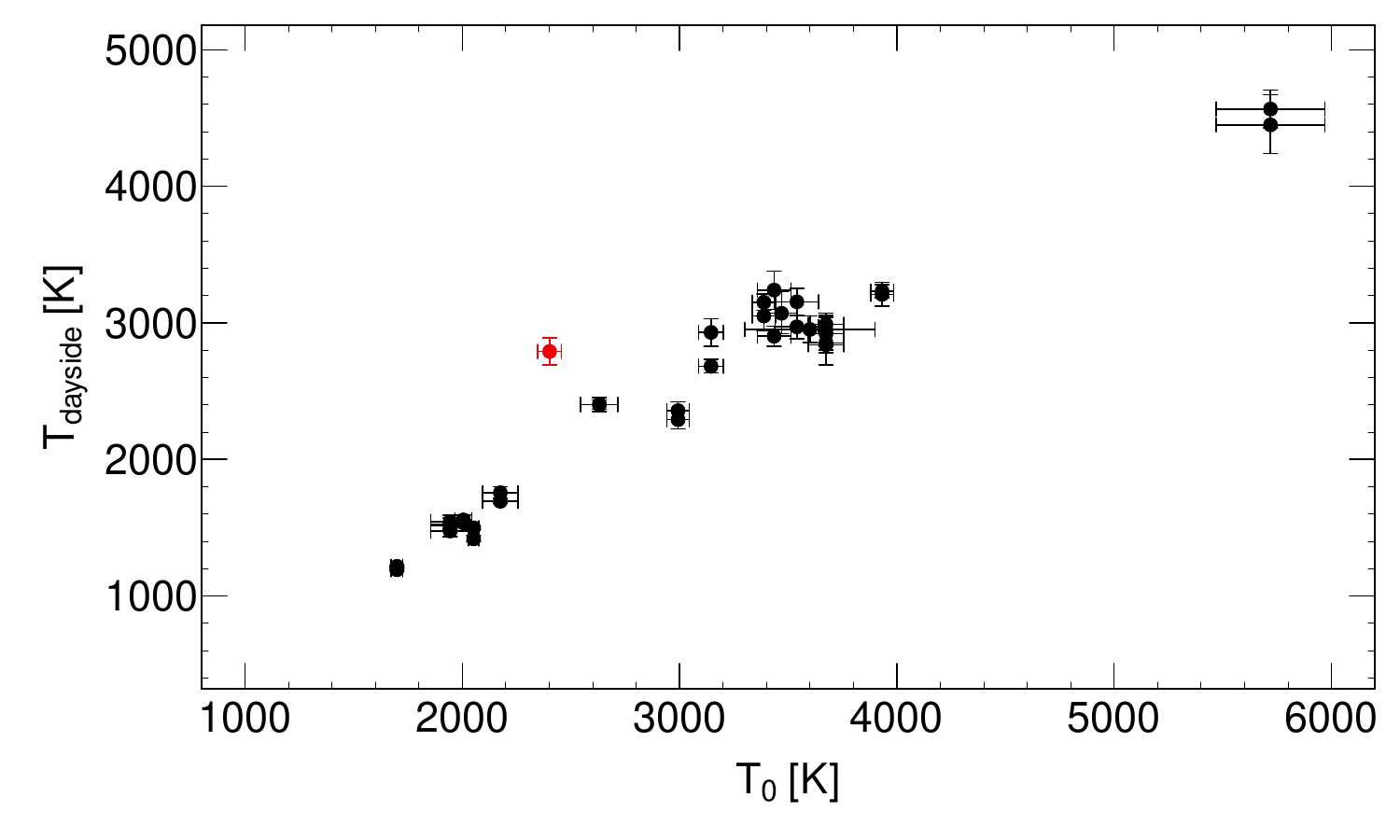}
    \caption{Dayside temperature as a function of equilibrium temperature from \cite{2021MNRAS.504.3316B} (black dots and error bars). Our result for\break WASP-167 is shown with red.}
    \label{fig:context-day}
\end{figure}

\begin{figure}
    \centering
    \includegraphics[width = \columnwidth]{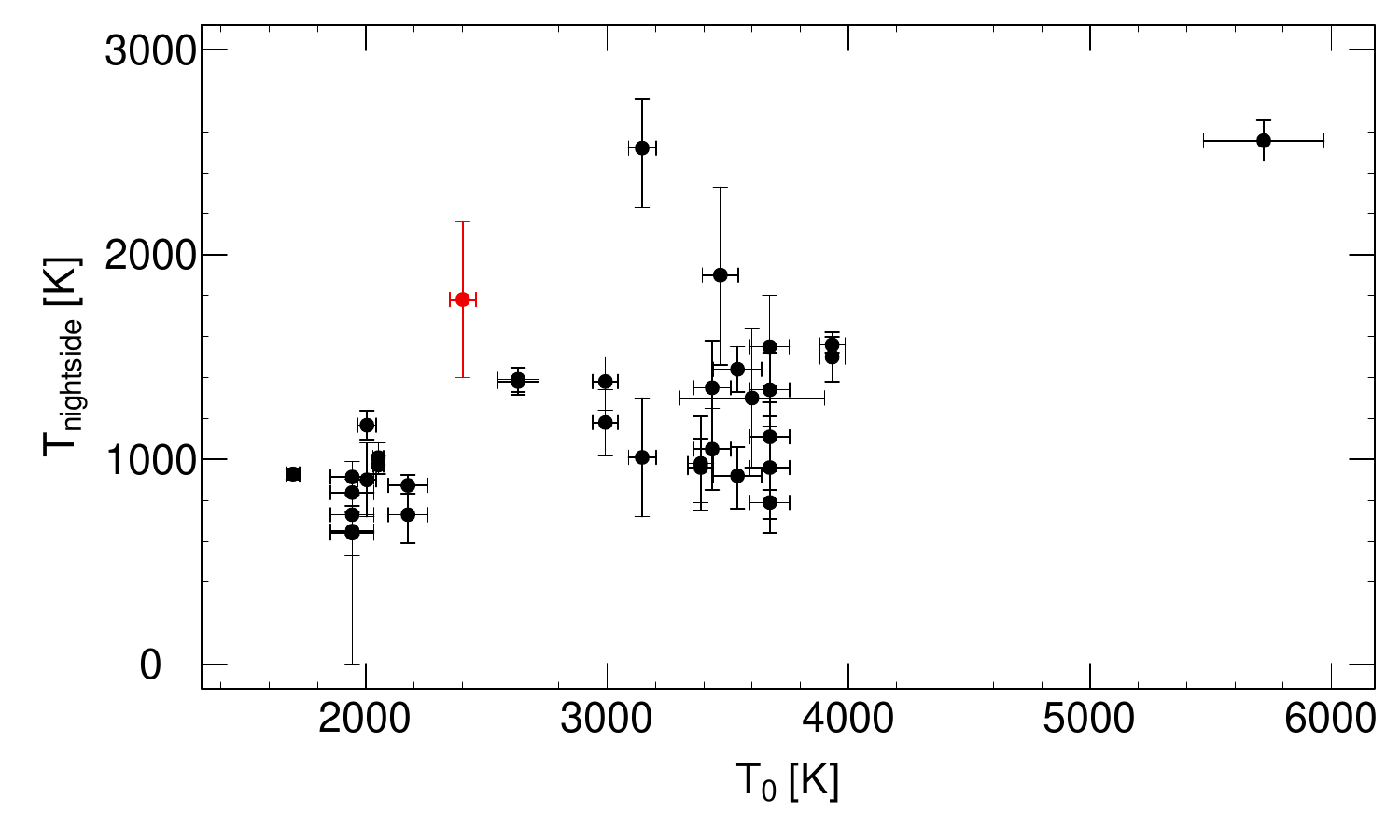}
    \caption{Nightside temperature as a function of equilibrium temperature from \cite{2021MNRAS.504.3316B} (black dots and error bars). Our result for\break WASP-167 is shown with red.}
    \label{fig:context-night}
\end{figure}

To put our results into context, we adopted the dayside, nightside and equilibrium temperatures, defined as ${T_0 = T_{\rm eff} \sqrt{\frac{R_\star}{2a}}}$, of several well studied exoplanets, from \cite{2021MNRAS.504.3316B}. We compare the derived $T_{\rm day}$ and $T_{\rm night}$ to these values on Figs. \ref{fig:context-day} and \ref{fig:context-night}. From these, we can conclude that our results are clear outliers from the planet population. Slight discrepancies might occur due to differences in the surface gravity on the planets or deviations from blackbody radiation. The latter could be enhanced as \cite{2021MNRAS.504.3316B} present results from infra red phase curves. We also note that the data adopted from \cite{2021MNRAS.504.3316B} is not compiled in a homogeneous way, and, as was presented in \cite{2022arXiv220801716K}, untreated time-correlated noise can induce biases in the modelling parameters. This would imply that such a direct comparison between different studies is not ideal. A further possibility for the observed deviation of our results from the overall population would be that we overestimate $T_{\rm day}$ and $T_{\rm night}$, by overestimating $T_{\rm intrinsic}$, which is related to $I_{\rm p}/I_\star$. The posterior distribution (Fig. \ref{fig:corner}) suggest that $I_{\rm p}/I_\star$ is compatible with $0$, and, by using the median of this distribution in our analysis, we might overestimate the amount of light emitted by the planet. A deeper exploration of the population statistics is beyond the scope of this paper.

Although the $T_{\rm intrinsic}$ of WASP-167b is poorly constrained, we explore the possible sources for the intrinsic heat. We list several possibilities in \cite{kelt9pow} and \cite{companion}, however, at the $1.29^{+0.36}_{-0.27}$~Gyr age of the system \citep{2017MNRAS.471.2743T}, with a planet on a circular orbit with a $\lambda = -167 \pm 5^\circ$ stellar obliquity, we assume that it originates from the so-called obliquity tides. \cite{2007A&A...462L...5L} and \cite{2019ApJ...886...72M} expressed the thermal contribution of this heating process as
\begin{equation}
    L_{\rm tide} = \frac{6\pi}{P} \frac{k_2}{Q_p} \frac{R_{\rm p}^5}{a^6} G M_\star^2,
\end{equation}
where $k_2$ and $Q_p$ are the Love number and tidal dissipation factor of the planet. The Love number can be measured by accounting for the slight changes in the shape of the transit \cite[e.g.][]{2019ApJ...878..119H, 2020ApJ...889...66H, 2022A&A...657A..52B, 2022A&A...658C...1B}, or, for eccentric orbits, via the modelling of the apsidal motion \citep{2019A&A...623A..45C}. A direct $k_2$-measurement for WASP-167b is beyond the scope of this paper, we therefore adopt $k_2 = 0.61$ from \cite{2019A&A...623A..45C}, a value which is also in line with $k_2$ measurements of Jupiter \citep[e.g.][]{2020GeoRL..4786572D}. We find that the $\sim 10^{-4}$~L$_\odot$ intrinsic luminosity of the planet yield $Q_p = (7.2 \pm 1.3)\cdot 10^5$, which is in line with \cite{kelt9pow} and the direct measurements for the giant planets of the Solar System \citep{1966Icar....5..375G}.

\section{Summary and conclusions}

We present the phase curve study of the WASP-167/KELT-13 system, discovered by \cite{2017MNRAS.471.2743T}, based on its \texttt{TESS} photometry from Sectors 10, 37 and 64. We confirm that the system hosts a hot Jupiter on a $\sim 2.022$ day orbit, and with a radius $\sim 1.6$ R$_J$ (Table \ref{sec:syspars}). We also present refined transit parameters. Previous attempts at constraining its mass from radial velocity data were unsuccessful due to the moderately rapid rotation ($52 \pm 8$~km s$^{-1}$), spectral type and activity of the host star. We found a peak in the Fourier spectrum of WASP-167 that is in near-resonance with the second orbital harmonic $\sim 0.989$~d$^{-1}$ (Fig. \ref{fig:fou}) of its companion -- the same time dependence as the ellipsoidal variability caused by WASP-167b. 

To analyse how the stellar pulsations affect the precision and accuracy of the transit parameters, we prepared 17 LCs, with an increasing number of sinusoidal components subtracted from them, according to Eq. \ref{eq:prep}. The parameters of these harmonics were obtained via the Fourier transform using \texttt{Period04}. The sets of examined out-of-transit parameters are shown in Figs.~\ref{fig:Kq} and \ref{fig:consts}. Using the Bayesian Information Criterion, we find that subtracting 16 sinusoidal curves yields the optimal solution. This is coincidentally the total number of harmonics that can be determined from the Fourier transform, with the commonly used ${\rm S/N} \geq 4$ criteria. This process of whitening the data yields self-consistent results, with the caveat that we may be overcompensating for the ellipsoidal variability. Thus, the obtained mass of WASP-167b at $0.34 \pm 0.22$~M$_J$ may be considerably underestimated. We also raise the possibility that the `critical frequency' influencing the mass estimates of WASP-167b may arise from planet-to-star interactions. If the stellar signal originates from pulsations, then the WASP-167 system shows similarities with HAT-P-2 \citep{2017ApJ...836L..17D}, or WASP-33 \citep{2022A&A...660L...2K} and HD 31221b \citep{2023arXiv230504000K}, depending on whether its orbit has a small (but non-zero) eccentricity or not. On the other hand, if the variability of WASP-167 originates from spots, then it is similar to AU Mic \citep{2021A&A...654A.159S} in this regard.

We also present the occultation of WASP-167b (Fig. \ref{fig:occ}), and its dependence on the stellar pulsations (Fig. \ref{fig:consts}), via modeling the phase curve of the planet. The resultant occultation depth of $106.8 \pm 27.3$ ppm, and the corresponding dayside temperature of $2790 \pm 100$ K is well below the limits provided by \cite{2017MNRAS.471.2743T} from ground-based data. It also raises the possibility that WASP-167b is an ultra-hot Jupiter, similar to WASP-33 \citep{2011MNRAS.416.2096S}. We measure a geometric albedo of $0.34 \pm 0.11$, and an eastward offset of the brightest point on the planetary surface of $17.2^\circ \pm 11.2^\circ$ -- similarly to WASP-33 in the same passband \citep{2020A&A...639A..34V}. Furthermore, we obtain ($2\sigma$) upper limits on the nightside and intrinsic temperature ($2360$~K and $2340$~K, respectively) and the Bond albedo ($0.51$), and discuss the possible combinations of Bond albedos and heat redistribution efficiencies (Fig. \ref{fig:AfT}) that can reproduce the observed phase curve (Fig. \ref{fig:occ}).

\begin{acknowledgements} 
This work was supported by the PRODEX Experiment Agreement No. 4000137122 between the ELTE E\"otv\"os Lor\'and University and the European Space Agency (ESA-D/SCI-LE-2021-0025). Support of the KKP-137523 `SeismoLab' \'Elvonal  grant  as well as the grant K-138962 of the Hungarian Research, Development and Innovation Office (NKFIH) are acknowledged. This project has been supported by the LP2021-9 Lend\" ulet grant of the Hungarian Academy of Sciences. Project no. C1746651 has been implemented with the support provided by the Ministry of Culture and Innovation of Hungary from the National Research, Development and Innovation Fund, financed under the NVKDP-2021 funding scheme. This project has received funding from the HUN-REN Hungarian Research Network under grant number KMP-2023/79. Supported by the DKOP-23 Doctoral Excellence Program of the Ministry for Culture and Innovation from the source of the National Research, Development and Innovation Fund.     
\end{acknowledgements}

 \bibliographystyle{aa}
\bibliography{refs}

\begin{thebibliography}{85}
\expandafter\ifx\csname natexlab\endcsname\relax\def\natexlab#1{#1}\fi

\bibitem[{{Aerts} {et~al.}(2018){Aerts}, {Molenberghs}, {Michielsen},
  {Pedersen}, {Bj{\"o}rklund}, {Johnston}, {Mombarg}, {Bowman}, {Buysschaert},
  {P{\'a}pics}, {Sekaran}, {Sundqvist}, {Tkachenko}, {Truyaert}, {Van Reeth},
  \& {Vermeyen}}]{2018ApJS..237...15A}
{Aerts}, C., {Molenberghs}, G., {Michielsen}, M., {et~al.} 2018, \apjs, 237, 15

\bibitem[{{Ahlers} {et~al.}(2019){Ahlers}, {Barnes}, \&
  {Myers}}]{2019AJ....158...88A}
{Ahlers}, J.~P., {Barnes}, J.~W., \& {Myers}, S.~A. 2019, \aj, 158, 88

\bibitem[{{Ahlers} {et~al.}(2022){Ahlers}, {Fromont}, {Kopparappu}, {Cauley},
  \& {Haqq-Misra}}]{2022ApJ...928...35A}
{Ahlers}, J.~P., {Fromont}, E.~F., {Kopparappu}, R., {Cauley}, P.~W., \&
  {Haqq-Misra}, J. 2022, \apj, 928, 35

\bibitem[{{Astropy Collaboration} {et~al.}(2022){Astropy Collaboration},
  {Price-Whelan}, {Lim}, {Earl}, {Starkman}, {Bradley}, {Shupe}, {Patil},
  {Corrales}, {Brasseur}, {N{\"o}the}, {Donath}, {Tollerud}, {Morris},
  {Ginsburg}, {Vaher}, {Weaver}, {Tocknell}, {Jamieson}, {van Kerkwijk},
  {Robitaille}, {Merry}, {Bachetti}, {G{\"u}nther}, {Aldcroft},
  {Alvarado-Montes}, {Archibald}, {B{\'o}di}, {Bapat}, {Barentsen},
  {Baz{\'a}n}, {Biswas}, {Boquien}, {Burke}, {Cara}, {Cara}, {Conroy},
  {Conseil}, {Craig}, {Cross}, {Cruz}, {D'Eugenio}, {Dencheva}, {Devillepoix},
  {Dietrich}, {Eigenbrot}, {Erben}, {Ferreira}, {Foreman-Mackey}, {Fox},
  {Freij}, {Garg}, {Geda}, {Glattly}, {Gondhalekar}, {Gordon}, {Grant},
  {Greenfield}, {Groener}, {Guest}, {Gurovich}, {Handberg}, {Hart},
  {Hatfield-Dodds}, {Homeier}, {Hosseinzadeh}, {Jenness}, {Jones}, {Joseph},
  {Kalmbach}, {Karamehmetoglu}, {Ka{\l}uszy{\'n}ski}, {Kelley}, {Kern},
  {Kerzendorf}, {Koch}, {Kulumani}, {Lee}, {Ly}, {Ma}, {MacBride}, {Maljaars},
  {Muna}, {Murphy}, {Norman}, {O'Steen}, {Oman}, {Pacifici}, {Pascual},
  {Pascual-Granado}, {Patil}, {Perren}, {Pickering}, {Rastogi}, {Roulston},
  {Ryan}, {Rykoff}, {Sabater}, {Sakurikar}, {Salgado}, {Sanghi}, {Saunders},
  {Savchenko}, {Schwardt}, {Seifert-Eckert}, {Shih}, {Jain}, {Shukla}, {Sick},
  {Simpson}, {Singanamalla}, {Singer}, {Singhal}, {Sinha}, {Sip{\H{o}}cz},
  {Spitler}, {Stansby}, {Streicher}, {{\v{S}}umak}, {Swinbank}, {Taranu},
  {Tewary}, {Tremblay}, {de Val-Borro}, {Van Kooten}, {Vasovi{\'c}}, {Verma},
  {de Miranda Cardoso}, {Williams}, {Wilson}, {Winkel}, {Wood-Vasey}, {Xue},
  {Yoachim}, {Zhang}, {Zonca}, \& {Astropy Project
  Contributors}}]{2022ApJ...935..167A}
{Astropy Collaboration}, {Price-Whelan}, A.~M., {Lim}, P.~L., {et~al.} 2022,
  \apj, 935, 167

\bibitem[{{Babcock}(1947)}]{1947ApJ...105..105B}
{Babcock}, H.~W. 1947, \apj, 105, 105

\bibitem[{{Bakos} {et~al.}(2007){Bakos}, {Kov{\'a}cs}, {Torres}, {Fischer},
  {Latham}, {Noyes}, {Sasselov}, {Mazeh}, {Shporer}, {Butler}, {Stefanik},
  {Fern{\'a}ndez}, {Sozzetti}, {P{\'a}l}, {Johnson}, {Marcy}, {Winn},
  {Sip{\H{o}}cz}, {L{\'a}z{\'a}r}, {Papp}, \& {S{\'a}ri}}]{2007ApJ...670..826B}
{Bakos}, G.~{\'A}., {Kov{\'a}cs}, G., {Torres}, G., {et~al.} 2007, \apj, 670,
  826

\bibitem[{{Barros} {et~al.}(2022{\natexlab{a}}){Barros}, {Akinsanmi},
  {Bou{\'e}}, {Smith}, {Laskar}, {Ulmer-Moll}, {Lillo-Box}, {Queloz},
  {Cameron}, {Sousa}, {Ehrenreich}, {Hooton}, {Bruno}, {Demory}, {Correia},
  {Demangeon}, {Wilson}, {Bonfanti}, {Hoyer}, {Alibert}, {Alonso},
  {Escud{\'e}}, {Barbato}, {B{\'a}rczy}, {Barrado}, {Baumjohann}, {Beck},
  {Beck}, {Benz}, {Bergomi}, {Billot}, {Bonfils}, {Bouchy}, {Brandeker},
  {Broeg}, {Cabrera}, {Cessa}, {Charnoz}, {Damme}, {Davies}, {Deleuil},
  {Deline}, {Delrez}, {Erikson}, {Fortier}, {Fossati}, {Fridlund}, {Gandolfi},
  {Mu{\~n}oz}, {Gillon}, {G{\"u}del}, {Isaak}, {Heng}, {Kiss}, {des Etangs},
  {Lendl}, {Lovis}, {Magrin}, {Nascimbeni}, {Maxted}, {Olofsson}, {Ottensamer},
  {Pagano}, {Pall{\'e}}, {Parviainen}, {Peter}, {Piotto}, {Pollacco},
  {Ragazzoni}, {Rando}, {Rauer}, {Ribas}, {Santos}, {Scandariato},
  {S{\'e}gransan}, {Simon}, {Steller}, {Szab{\'o}}, {Thomas}, {Udry}, {Ulmer},
  {Van Grootel}, \& {Walton}}]{2022A&A...657A..52B}
{Barros}, S.~C.~C., {Akinsanmi}, B., {Bou{\'e}}, G., {et~al.}
  2022{\natexlab{a}}, \aap, 657, A52

\bibitem[{{Barros} {et~al.}(2022{\natexlab{b}}){Barros}, {Akinsanmi},
  {Bou{\'e}}, {Smith}, {Laskar}, {Ulmer-Moll}, {Lillo-Box}, {Queloz}, {Collier
  Cameron}, {Sousa}, {Ehrenreich}, {Hooton}, {Bruno}, {Demory}, {Correia},
  {Demangeon}, {Wilson}, {Bonfanti}, {Hoyer}, {Alibert}, {Alonso}, {Anglada
  Escud{\'e}}, {Barbato}, {B{\'a}rczy}, {Barrado}, {Baumjohann}, {Beck},
  {Beck}, {Benz}, {Bergomi}, {Billot}, {Bonfils}, {Bouchy}, {Brandeker},
  {Broeg}, {Cabrera}, {Cessa}, {Charnoz}, {Damme}, {Davies}, {Deleuil},
  {Deline}, {Delrez}, {Erikson}, {Fortier}, {Fossati}, {Fridlund}, {Gandolfi},
  {Garc{\'\i}a Mu{\~n}oz}, {Gillon}, {G{\"u}del}, {Isaak}, {Heng}, {Kiss},
  {Lecavelier des Etangs}, {Lendl}, {Lovis}, {Magrin}, {Nascimbeni}, {Maxted},
  {Olofsson}, {Ottensamer}, {Pagano}, {Pall{\'e}}, {Parviainen}, {Peter},
  {Piotto}, {Pollacco}, {Ragazzoni}, {Rando}, {Rauer}, {Ribas}, {Salmon},
  {Santos}, {Scandariato}, {S{\'e}gransan}, {Simon}, {Steller}, {Szab{\'o}},
  {Thomas}, {Udry}, {Ulmer}, {Van Grootel}, \& {Walton}}]{2022A&A...658C...1B}
{Barros}, S.~C.~C., {Akinsanmi}, B., {Bou{\'e}}, G., {et~al.}
  2022{\natexlab{b}}, \aap, 658, C1

\bibitem[{{Bell} {et~al.}(2021){Bell}, {Dang}, {Cowan}, {Bean}, {D{\'e}sert},
  {Fortney}, {Keating}, {Kempton}, {Kreidberg}, {Line}, {Mansfield},
  {Parmentier}, {Stevenson}, {Swain}, \& {Zellem}}]{2021MNRAS.504.3316B}
{Bell}, T.~J., {Dang}, L., {Cowan}, N.~B., {et~al.} 2021, \mnras, 504, 3316

\bibitem[{{B{\'o}kon} {et~al.}(2023){B{\'o}kon}, {K{\'a}lm{\'a}n},
  {B{\'\i}r{\'o}}, \& {Szab{\'o}}}]{2023A&A...674A.186B}
{B{\'o}kon}, A., {K{\'a}lm{\'a}n}, S., {B{\'\i}r{\'o}}, I.~B., \& {Szab{\'o}},
  M.~G. 2023, \aap, 674, A186

\bibitem[{{Borucki} {et~al.}(2010){Borucki}, {Koch}, {Basri}, {Batalha},
  {Brown}, {Caldwell}, {Caldwell}, {Christensen-Dalsgaard}, {Cochran},
  {DeVore}, {Dunham}, {Dupree}, {Gautier}, {Geary}, {Gilliland}, {Gould},
  {Howell}, {Jenkins}, {Kondo}, {Latham}, {Marcy}, {Meibom}, {Kjeldsen},
  {Lissauer}, {Monet}, {Morrison}, {Sasselov}, {Tarter}, {Boss}, {Brownlee},
  {Owen}, {Buzasi}, {Charbonneau}, {Doyle}, {Fortney}, {Ford}, {Holman},
  {Seager}, {Steffen}, {Welsh}, {Rowe}, {Anderson}, {Buchhave}, {Ciardi},
  {Walkowicz}, {Sherry}, {Horch}, {Isaacson}, {Everett}, {Fischer}, {Torres},
  {Johnson}, {Endl}, {MacQueen}, {Bryson}, {Dotson}, {Haas}, {Kolodziejczak},
  {Van Cleve}, {Chandrasekaran}, {Twicken}, {Quintana}, {Clarke}, {Allen},
  {Li}, {Wu}, {Tenenbaum}, {Verner}, {Bruhweiler}, {Barnes}, \&
  {Prsa}}]{2010Sci...327..977B}
{Borucki}, W.~J., {Koch}, D., {Basri}, G., {et~al.} 2010, Science, 327, 977

\bibitem[{{Bowman} {et~al.}(2020){Bowman}, {Burssens}, {Sim{\'o}n-D{\'\i}az},
  {Edelmann}, {Rogers}, {Horst}, {R{\"o}pke}, \& {Aerts}}]{2020A&A...640A..36B}
{Bowman}, D.~M., {Burssens}, S., {Sim{\'o}n-D{\'\i}az}, S., {et~al.} 2020,
  \aap, 640, A36

\bibitem[{{Bravo} {et~al.}(2014){Bravo}, {Roque}, {Estrela}, {Le{\~a}o}, \& {De
  Medeiros}}]{2014A&A...568A..34B}
{Bravo}, J.~P., {Roque}, S., {Estrela}, R., {Le{\~a}o}, I.~C., \& {De
  Medeiros}, J.~R. 2014, \aap, 568, A34

\bibitem[{{Breger} {et~al.}(1993){Breger}, {Stich}, {Garrido}, {Martin},
  {Jiang}, {Li}, {Hube}, {Ostermann}, {Paparo}, \&
  {Scheck}}]{1993A&A...271..482B}
{Breger}, M., {Stich}, J., {Garrido}, R., {et~al.} 1993, \aap, 271, 482

\bibitem[{{Bryan} {et~al.}(2024){Bryan}, {de Wit}, {Sun}, {de Beurs}, \&
  {Townsend}}]{2024arXiv240308014B}
{Bryan}, J., {de Wit}, J., {Sun}, M., {de Beurs}, Z.~L., \& {Townsend}, R.
  H.~D. 2024, arXiv e-prints, arXiv:2403.08014

\bibitem[{{Carter} \& {Winn}(2009)}]{2009ApJ...704...51C}
{Carter}, J.~A. \& {Winn}, J.~N. 2009, \apj, 704, 51

\bibitem[{{Christian} {et~al.}(2006){Christian}, {Pollacco}, {Skillen},
  {Street}, {Keenan}, {Clarkson}, {Collier Cameron}, {Kane}, {Lister}, {West},
  {Enoch}, {Evans}, {Fitzsimmons}, {Haswell}, {Hellier}, {Hodgkin}, {Horne},
  {Irwin}, {Norton}, {Osborne}, {Ryans}, {Wheatley}, \&
  {Wilson}}]{2006MNRAS.372.1117C}
{Christian}, D.~J., {Pollacco}, D.~L., {Skillen}, I., {et~al.} 2006, \mnras,
  372, 1117

\bibitem[{{Claret}(2000)}]{2000A&A...363.1081C}
{Claret}, A. 2000, \aap, 363, 1081

\bibitem[{{Claret}(2004)}]{2004A&A...428.1001C}
{Claret}, A. 2004, \aap, 428, 1001

\bibitem[{{Collier Cameron} {et~al.}(2010){Collier Cameron}, {Guenther},
  {Smalley}, {McDonald}, {Hebb}, {Andersen}, {Augusteijn}, {Barros}, {Brown},
  {Cochran}, {Endl}, {Fossey}, {Hartmann}, {Maxted}, {Pollacco}, {Skillen},
  {Telting}, {Waldmann}, \& {West}}]{2010MNRAS.407..507C}
{Collier Cameron}, A., {Guenther}, E., {Smalley}, B., {et~al.} 2010, \mnras,
  407, 507

\bibitem[{{Cowan} \& {Agol}(2011)}]{2011ApJ...729...54C}
{Cowan}, N.~B. \& {Agol}, E. 2011, \apj, 729, 54

\bibitem[{{Csizmadia}(2020)}]{2020MNRAS.496.4442C}
{Csizmadia}, S. 2020, \mnras, 496, 4442

\bibitem[{{Csizmadia} {et~al.}(2019){Csizmadia}, {Hellard}, \&
  {Smith}}]{2019A&A...623A..45C}
{Csizmadia}, S., {Hellard}, H., \& {Smith}, A.~M.~S. 2019, \aap, 623, A45

\bibitem[{{Csizmadia} {et~al.}(2023{\natexlab{a}}){Csizmadia}, {Smith},
  {Cabrera}, {Klagyivik}, {Chaushev}, \& {Lam}}]{kelt9pow}
{Csizmadia}, S., {Smith}, A.~M.~S., {Cabrera}, J., {et~al.} 2023{\natexlab{a}},
  \aap, submitted

\bibitem[{{Csizmadia} {et~al.}(2023{\natexlab{b}}){Csizmadia}, {Smith},
  {K{\'a}lm{\'a}n}, {Cabrera}, {Klagyivik}, {Chaushev}, \&
  {Lam}}]{2021arXiv210811822C}
{Csizmadia}, S., {Smith}, A.~M.~S., {K{\'a}lm{\'a}n}, S., {et~al.}
  2023{\natexlab{b}}, \aap, 675, A106

\bibitem[{{de Wit} {et~al.}(2017){de Wit}, {Lewis}, {Knutson}, {Fuller},
  {Antoci}, {Fulton}, {Laughlin}, {Deming}, {Shporer}, {Batygin}, {Cowan},
  {Agol}, {Burrows}, {Fortney}, {Langton}, \& {Showman}}]{2017ApJ...836L..17D}
{de Wit}, J., {Lewis}, N.~K., {Knutson}, H.~A., {et~al.} 2017, \apjl, 836, L17

\bibitem[{{Deming} {et~al.}(2012){Deming}, {Fraine}, {Sada}, {Madhusudhan},
  {Knutson}, {Harrington}, {Blecic}, {Nymeyer}, {Smith}, \&
  {Jackson}}]{2012ApJ...754..106D}
{Deming}, D., {Fraine}, J.~D., {Sada}, P.~V., {et~al.} 2012, \apj, 754, 106

\bibitem[{{Deming} \& {Knutson}(2020)}]{2020NatAs...4..453D}
{Deming}, D. \& {Knutson}, H.~A. 2020, Nature Astronomy, 4, 453

\bibitem[{{Derekas} {et~al.}(2009){Derekas}, {Kiss}, {Bedding}, {Ashley},
  {Cs{\'a}k}, {Danos}, {Fernandez}, {F{\H{u}}r{\'e}sz}, {M{\'e}sz{\'a}ros},
  {Szab{\'o}}, {Szak{\'a}ts}, {Sz{\'e}kely}, \&
  {Szatm{\'a}ry}}]{2009MNRAS.394..995D}
{Derekas}, A., {Kiss}, L.~L., {Bedding}, T.~R., {et~al.} 2009, \mnras, 394, 995

\bibitem[{{Derekas} {et~al.}(2017){Derekas}, {Plachy}, {Moln{\'a}r},
  {S{\'o}dor}, {Benk{\H{o}}}, {Szabados}, {Bogn{\'a}r}, {Cs{\'a}k},
  {Szab{\'o}}, {Szab{\'o}}, \& {P{\'a}l}}]{2017MNRAS.464.1553D}
{Derekas}, A., {Plachy}, E., {Moln{\'a}r}, L., {et~al.} 2017, \mnras, 464, 1553

\bibitem[{{Durante} {et~al.}(2020){Durante}, {Parisi}, {Serra}, {Zannoni},
  {Notaro}, {Racioppa}, {Buccino}, {Lari}, {Gomez Casajus}, {Iess}, {Folkner},
  {Tommei}, {Tortora}, \& {Bolton}}]{2020GeoRL..4786572D}
{Durante}, D., {Parisi}, M., {Serra}, D., {et~al.} 2020, \grl, 47, e86572

\bibitem[{{Faigler} \& {Mazeh}(2011)}]{2011MNRAS.415.3921F}
{Faigler}, S. \& {Mazeh}, T. 2011, \mnras, 415, 3921

\bibitem[{{Foreman-Mackey} {et~al.}(2017){Foreman-Mackey}, {Agol},
  {Ambikasaran}, \& {Angus}}]{2017AJ....154..220F}
{Foreman-Mackey}, D., {Agol}, E., {Ambikasaran}, S., \& {Angus}, R. 2017, \aj,
  154, 220

\bibitem[{{Foster}(1996)}]{1996AJ....112.1709F}
{Foster}, G. 1996, \aj, 112, 1709

\bibitem[{{Ginsburg} {et~al.}(2019){Ginsburg}, {Sip{\H o}cz}, {Brasseur},
  {Cowperthwaite}, {Craig}, {Deil}, {Guillochon}, {Guzman}, {Liedtke}, {Lian
  Lim}, {Lockhart}, {Mommert}, {Morris}, {Norman}, {Parikh}, {Persson},
  {Robitaille}, {Segovia}, {Singer}, {Tollerud}, {de Val-Borro}, {Valtchanov},
  {Woillez}, {The Astroquery collaboration}, \& {a subset of the astropy
  collaboration}}]{2019AJ....157...98G}
{Ginsburg}, A., {Sip{\H o}cz}, B.~M., {Brasseur}, C.~E., {et~al.} 2019, \aj,
  157, 98

\bibitem[{{Goldreich} \& {Soter}(1966)}]{1966Icar....5..375G}
{Goldreich}, P. \& {Soter}, S. 1966, \icarus, 5, 375

\bibitem[{{Hartman} {et~al.}(2015){Hartman}, {Bakos}, {Buchhave}, {Torres},
  {Latham}, {Kov{\'a}cs}, {Bhatti}, {Csubry}, {de Val-Borro}, {Penev}, {Huang},
  {B{\'e}ky}, {Bieryla}, {Quinn}, {Howard}, {Marcy}, {Johnson}, {Isaacson},
  {Fischer}, {Noyes}, {Falco}, {Esquerdo}, {Knox}, {Hinz}, {L{\'a}z{\'a}r},
  {Papp}, \& {S{\'a}ri}}]{2015AJ....150..197H}
{Hartman}, J.~D., {Bakos}, G.~{\'A}., {Buchhave}, L.~A., {et~al.} 2015, \aj,
  150, 197

\bibitem[{{Haynes} {et~al.}(2015){Haynes}, {Mandell}, {Madhusudhan}, {Deming},
  \& {Knutson}}]{2015ApJ...806..146H}
{Haynes}, K., {Mandell}, A.~M., {Madhusudhan}, N., {Deming}, D., \& {Knutson},
  H. 2015, \apj, 806, 146

\bibitem[{{Hellard} {et~al.}(2019){Hellard}, {Csizmadia}, {Padovan}, {Rauer},
  {Cabrera}, {Sohl}, {Spohn}, \& {Breuer}}]{2019ApJ...878..119H}
{Hellard}, H., {Csizmadia}, S., {Padovan}, S., {et~al.} 2019, \apj, 878, 119

\bibitem[{{Hellard} {et~al.}(2020){Hellard}, {Csizmadia}, {Padovan}, {Sohl}, \&
  {Rauer}}]{2020ApJ...889...66H}
{Hellard}, H., {Csizmadia}, S., {Padovan}, S., {Sohl}, F., \& {Rauer}, H. 2020,
  \apj, 889, 66

\bibitem[{{Herman} {et~al.}(2018){Herman}, {de Mooij}, {Huang}, \&
  {Jayawardhana}}]{2018AJ....155...13H}
{Herman}, M.~K., {de Mooij}, E. J.~W., {Huang}, C.~X., \& {Jayawardhana}, R.
  2018, \aj, 155, 13

\bibitem[{{Herrero} {et~al.}(2011){Herrero}, {Morales}, {Ribas}, \&
  {Naves}}]{2011A&A...526L..10H}
{Herrero}, E., {Morales}, J.~C., {Ribas}, I., \& {Naves}, R. 2011, \aap, 526,
  L10

\bibitem[{{Hey} {et~al.}(2021){Hey}, {Montet}, {Pope}, {Murphy}, \&
  {Bedding}}]{2021AJ....162..204H}
{Hey}, D.~R., {Montet}, B.~T., {Pope}, B. J.~S., {Murphy}, S.~J., \& {Bedding},
  T.~R. 2021, \aj, 162, 204

\bibitem[{{Hurley} {et~al.}(2000){Hurley}, {Pols}, \&
  {Tout}}]{2000MNRAS.315..543H}
{Hurley}, J.~R., {Pols}, O.~R., \& {Tout}, C.~A. 2000, \mnras, 315, 543

\bibitem[{Husser {et~al.}(2013)Husser, {Wende-von Berg}, Dreizler, Homeier,
  Reiners, Barman, \& Hauschildt}]{Husser2013}
Husser, T.-O., {Wende-von Berg}, S., Dreizler, S., {et~al.} 2013, A{\&}A, 553,
  A6

\bibitem[{{Johnson} {et~al.}(2015){Johnson}, {Cochran}, {Collier Cameron}, \&
  {Bayliss}}]{2015ApJ...810L..23J}
{Johnson}, M.~C., {Cochran}, W.~D., {Collier Cameron}, A., \& {Bayliss}, D.
  2015, \apjl, 810, L23

\bibitem[{{K{\'a}lm{\'a}n} {et~al.}(2022){K{\'a}lm{\'a}n}, {B{\'o}kon},
  {Derekas}, {Szab{\'o}}, {Heged{\H{u}}s}, \& {Nagy}}]{2022A&A...660L...2K}
{K{\'a}lm{\'a}n}, S., {B{\'o}kon}, A., {Derekas}, A., {et~al.} 2022, \aap, 660,
  L2

\bibitem[{{K{\'a}lm{\'a}n} {et~al.}(2024){K{\'a}lm{\'a}n}, {Csizmadia},
  Derekas, Heged\H{u}s, P\'al, Szab\'o, \& M.}]{companion}
{K{\'a}lm{\'a}n}, S., {Csizmadia}, S., Derekas, A., {et~al.} 2024, \aj,
  submitted

\bibitem[{{K{\'a}lm{\'a}n} {et~al.}(2023{\natexlab{a}}){K{\'a}lm{\'a}n},
  {Derekas}, {Csizmadia}, {Szab{\'o}}, {Heged{\H{u}}s}, {Smith}, {Kov{\'a}cs},
  {Ziegler}, {P{\'a}l}, {Szab{\'o}}, {Parviainen}, \&
  {Murgas}}]{2023arXiv230504000K}
{K{\'a}lm{\'a}n}, S., {Derekas}, A., {Csizmadia}, S., {et~al.}
  2023{\natexlab{a}}, \aap, 673, L14

\bibitem[{{K{\'a}lm{\'a}n} {et~al.}(2023{\natexlab{b}}){K{\'a}lm{\'a}n},
  {Szab{\'o}}, \& {Csizmadia}}]{2022arXiv220801716K}
{K{\'a}lm{\'a}n}, S., {Szab{\'o}}, G.~M., \& {Csizmadia}, S.
  2023{\natexlab{b}}, \aap, 675, A107

\bibitem[{{Knutson} {et~al.}(2009){Knutson}, {Charbonneau}, {Cowan}, {Fortney},
  {Showman}, {Agol}, \& {Henry}}]{2009ApJ...703..769K}
{Knutson}, H.~A., {Charbonneau}, D., {Cowan}, N.~B., {et~al.} 2009, \apj, 703,
  769

\bibitem[{{Kov{\'a}cs} {et~al.}(2013){Kov{\'a}cs}, {Kov{\'a}cs}, {Hartman},
  {Bakos}, {Bieryla}, {Latham}, {Noyes}, {Reg{\'a}ly}, \&
  {Esquerdo}}]{2013A&A...553A..44K}
{Kov{\'a}cs}, G., {Kov{\'a}cs}, T., {Hartman}, J.~D., {et~al.} 2013, \aap, 553,
  A44

\bibitem[{{Kurtz} {et~al.}(2015){Kurtz}, {Shibahashi}, {Murphy}, {Bedding}, \&
  {Bowman}}]{2015MNRAS.450.3015K}
{Kurtz}, D.~W., {Shibahashi}, H., {Murphy}, S.~J., {Bedding}, T.~R., \&
  {Bowman}, D.~M. 2015, \mnras, 450, 3015

\bibitem[{{Lehmann} {et~al.}(2015){Lehmann}, {Guenther}, {Sebastian},
  {D{\"o}llinger}, {Hartmann}, \& {Mkrtichian}}]{2015A&A...578L...4L}
{Lehmann}, H., {Guenther}, E., {Sebastian}, D., {et~al.} 2015, \aap, 578, L4

\bibitem[{{Lenz} \& {Breger}(2005)}]{2005CoAst.146...53L}
{Lenz}, P. \& {Breger}, M. 2005, Communications in Asteroseismology, 146, 53

\bibitem[{{Levrard} {et~al.}(2007){Levrard}, {Correia}, {Chabrier}, {Baraffe},
  {Selsis}, \& {Laskar}}]{2007A&A...462L...5L}
{Levrard}, B., {Correia}, A.~C.~M., {Chabrier}, G., {et~al.} 2007, \aap, 462,
  L5

\bibitem[{{Lightkurve Collaboration} {et~al.}(2018){Lightkurve Collaboration},
  {Cardoso}, {Hedges}, {Gully-Santiago}, {Saunders}, {Cody}, {Barclay}, {Hall},
  {Sagear}, {Turtelboom}, {Zhang}, {Tzanidakis}, {Mighell}, {Coughlin}, {Bell},
  {Berta-Thompson}, {Williams}, {Dotson}, \& {Barentsen}}]{2018ascl.soft12013L}
{Lightkurve Collaboration}, {Cardoso}, J.~V.~d.~M., {Hedges}, C., {et~al.}
  2018, {Lightkurve: Kepler and TESS time series analysis in Python},
  Astrophysics Source Code Library

\bibitem[{{Mandel} \& {Agol}(2002)}]{2002ApJ...580L.171M}
{Mandel}, K. \& {Agol}, E. 2002, \apjl, 580, L171

\bibitem[{{Mazeh} {et~al.}(2012){Mazeh}, {Nachmani}, {Sokol}, {Faigler}, \&
  {Zucker}}]{2012A&A...541A..56M}
{Mazeh}, T., {Nachmani}, G., {Sokol}, G., {Faigler}, S., \& {Zucker}, S. 2012,
  \aap, 541, A56

\bibitem[{{Millholland}(2019)}]{2019ApJ...886...72M}
{Millholland}, S. 2019, \apj, 886, 72

\bibitem[{{Nymeyer} {et~al.}(2011){Nymeyer}, {Harrington}, {Hardy},
  {Stevenson}, {Campo}, {Madhusudhan}, {Collier-Cameron}, {Loredo}, {Blecic},
  {Bowman}, {Britt}, {Cubillos}, {Hellier}, {Gillon}, {Maxted}, {Hebb},
  {Wheatley}, {Pollacco}, \& {Anderson}}]{2011ApJ...742...35N}
{Nymeyer}, S., {Harrington}, J., {Hardy}, R.~A., {et~al.} 2011, \apj, 742, 35

\bibitem[{Parviainen \& Aigrain(2015)}]{Parviainen2015}
Parviainen, H. \& Aigrain, S. 2015, MNRAS, 453, 3821

\bibitem[{{Rauer} {et~al.}(2014){Rauer}, {Catala}, {Aerts}, {Appourchaux},
  {Benz}, {Brandeker}, {Christensen-Dalsgaard}, {Deleuil}, {Gizon}, {Goupil},
  {G{\"u}del}, {Janot-Pacheco}, {Mas-Hesse}, {Pagano}, {Piotto}, {Pollacco},
  {Santos}, {Smith}, {Su{\'a}rez}, {Szab{\'o}}, {Udry}, {Adibekyan}, {Alibert},
  {Almenara}, {Amaro-Seoane}, {Eiff}, {Asplund}, {Antonello}, {Barnes},
  {Baudin}, {Belkacem}, {Bergemann}, {Bihain}, {Birch}, {Bonfils}, {Boisse},
  {Bonomo}, {Borsa}, {Brand{\~a}o}, {Brocato}, {Brun}, {Burleigh}, {Burston},
  {Cabrera}, {Cassisi}, {Chaplin}, {Charpinet}, {Chiappini}, {Church},
  {Csizmadia}, {Cunha}, {Damasso}, {Davies}, {Deeg}, {D{\'\i}az}, {Dreizler},
  {Dreyer}, {Eggenberger}, {Ehrenreich}, {Eigm{\"u}ller}, {Erikson}, {Farmer},
  {Feltzing}, {de Oliveira Fialho}, {Figueira}, {Forveille}, {Fridlund},
  {Garc{\'\i}a}, {Giommi}, {Giuffrida}, {Godolt}, {Gomes da Silva}, {Granzer},
  {Grenfell}, {Grotsch-Noels}, {G{\"u}nther}, {Haswell}, {Hatzes},
  {H{\'e}brard}, {Hekker}, {Helled}, {Heng}, {Jenkins}, {Johansen},
  {Khodachenko}, {Kislyakova}, {Kley}, {Kolb}, {Krivova}, {Kupka}, {Lammer},
  {Lanza}, {Lebreton}, {Magrin}, {Marcos-Arenal}, {Marrese}, {Marques},
  {Martins}, {Mathis}, {Mathur}, {Messina}, {Miglio}, {Montalban}, {Montalto},
  {Monteiro}, {Moradi}, {Moravveji}, {Mordasini}, {Morel}, {Mortier},
  {Nascimbeni}, {Nelson}, {Nielsen}, {Noack}, {Norton}, {Ofir}, {Oshagh},
  {Ouazzani}, {P{\'a}pics}, {Parro}, {Petit}, {Plez}, {Poretti}, {Quirrenbach},
  {Ragazzoni}, {Raimondo}, {Rainer}, {Reese}, {Redmer}, {Reffert},
  {Rojas-Ayala}, {Roxburgh}, {Salmon}, {Santerne}, {Schneider}, {Schou},
  {Schuh}, {Schunker}, {Silva-Valio}, {Silvotti}, {Skillen}, {Snellen}, {Sohl},
  {Sousa}, {Sozzetti}, {Stello}, {Strassmeier}, {{\v{S}}vanda}, {Szab{\'o}},
  {Tkachenko}, {Valencia}, {Van Grootel}, {Vauclair}, {Ventura}, {Wagner},
  {Walton}, {Weingrill}, {Werner}, {Wheatley}, \&
  {Zwintz}}]{2014ExA....38..249R}
{Rauer}, H., {Catala}, C., {Aerts}, C., {et~al.} 2014, Experimental Astronomy,
  38, 249

\bibitem[{{Ricker} {et~al.}(2015){Ricker}, {Winn}, {Vanderspek}, {Latham},
  {Bakos}, {Bean}, {Berta-Thompson}, {Brown}, {Buchhave}, {Butler}, {Butler},
  {Chaplin}, {Charbonneau}, {Christensen-Dalsgaard}, {Clampin}, {Deming},
  {Doty}, {De Lee}, {Dressing}, {Dunham}, {Endl}, {Fressin}, {Ge}, {Henning},
  {Holman}, {Howard}, {Ida}, {Jenkins}, {Jernigan}, {Johnson}, {Kaltenegger},
  {Kawai}, {Kjeldsen}, {Laughlin}, {Levine}, {Lin}, {Lissauer}, {MacQueen},
  {Marcy}, {McCullough}, {Morton}, {Narita}, {Paegert}, {Palle}, {Pepe},
  {Pepper}, {Quirrenbach}, {Rinehart}, {Sasselov}, {Sato}, {Seager},
  {Sozzetti}, {Stassun}, {Sullivan}, {Szentgyorgyi}, {Torres}, {Udry}, \&
  {Villasenor}}]{2015JATIS...1a4003R}
{Ricker}, G.~R., {Winn}, J.~N., {Vanderspek}, R., {et~al.} 2015, Journal of
  Astronomical Telescopes, Instruments, and Systems, 1, 014003

\bibitem[{{Rowe} {et~al.}(2008){Rowe}, {Matthews}, {Seager}, {Miller-Ricci},
  {Sasselov}, {Kuschnig}, {Guenther}, {Moffat}, {Rucinski}, {Walker}, \&
  {Weiss}}]{2008ApJ...689.1345R}
{Rowe}, J.~F., {Matthews}, J.~M., {Seager}, S., {et~al.} 2008, \apj, 689, 1345

\bibitem[{{Samadi-Ghadim} {et~al.}(2022){Samadi-Ghadim}, {Lampens}, \&
  {Gizon}}]{2022A&A...667A..60S}
{Samadi-Ghadim}, A., {Lampens}, P., \& {Gizon}, L. 2022, \aap, 667, A60

\bibitem[{{Singh} {et~al.}(2023){Singh}, {Scandariato}, {Smith}, {Cubillos},
  {Lendl}, {Billot}, {Fortier}, {Queloz}, {Sousa}, {Csizmadia}, {Brandeker},
  {Carone}, {Wilson}, {Akinsanmi}, {Patel}, {Krenn}, {Demangeon}, {Bruno},
  {Pagano}, {Hooton}, {Cabrera}, {Santos}, {Alibert}, {Alonso}, {Asquier},
  {B{\'a}rczy}, {Barrado Navascues}, {Barros}, {Baumjohann}, {Beck}, {Beck},
  {Benz}, {Bergomi}, {Bonfanti}, {Bonfils}, {Borsato}, {Broeg}, {Charnoz},
  {Collier Cameron}, {Davies}, {Deleuil}, {Deline}, {Delrez}, {Demory},
  {Ehrenreich}, {Erikson}, {Fossati}, {Fridlund}, {Gandolfi}, {Gillon},
  {G{\"u}del}, {G{\"u}nther}, {Harre}, {Heitzmann}, {Helling}, {Hoyer},
  {Isaak}, {Kiss}, {Lam}, {Laskar}, {Lecavelier des Etangs}, {Magrin},
  {Maxted}, {Mischler}, {Mordasini}, {Nascimbeni}, {Olofsson}, {Ottensamer},
  {Pall{\'e}}, {Peter}, {Piotto}, {Pollacco}, {Ragazzoni}, {Rando}, {Rauer},
  {Ribas}, {Salmon}, {S{\'e}gransan}, {Simon}, {Stalport}, {Steinberger},
  {Szab{\'o}}, {Thomas}, {Udry}, {Ulmer}, {Van Grootel}, {Venturini},
  {Villaver}, {Walton}, \& {Zingales}}]{2023arXiv231103264S}
{Singh}, V., {Scandariato}, G., {Smith}, A.~M.~S., {et~al.} 2023, arXiv
  e-prints, arXiv:2311.03264

\bibitem[{{Smith} {et~al.}(2011){Smith}, {Anderson}, {Skillen}, {Collier
  Cameron}, \& {Smalley}}]{2011MNRAS.416.2096S}
{Smith}, A.~M.~S., {Anderson}, D.~R., {Skillen}, I., {Collier Cameron}, A., \&
  {Smalley}, B. 2011, \mnras, 416, 2096

\bibitem[{{Southworth} {et~al.}(2020){Southworth}, {Bowman}, {Tkachenko}, \&
  {Pavlovski}}]{2020MNRAS.497L..19S}
{Southworth}, J., {Bowman}, D.~M., {Tkachenko}, A., \& {Pavlovski}, K. 2020,
  \mnras, 497, L19

\bibitem[{{Sowicka} {et~al.}(2023){Sowicka}, {Handler}, {Jones}, {Caldwell},
  {van Wyk}, {Paunzen}, {B{\k{a}}kowska}, {Peralta de Arriba},
  {Su{\'a}rez-Andr{\'e}s}, {Werner}, {Karjalainen}, \&
  {Holdsworth}}]{2023ApJS..269...32S}
{Sowicka}, P., {Handler}, G., {Jones}, D., {et~al.} 2023, \apjs, 269, 32

\bibitem[{{Steindl} {et~al.}(2021){Steindl}, {Zwintz}, \&
  {Bowman}}]{2021A&A...645A.119S}
{Steindl}, T., {Zwintz}, K., \& {Bowman}, D.~M. 2021, \aap, 645, A119

\bibitem[{{Stephan} {et~al.}(2022){Stephan}, {Wang}, {Cauley}, {Gaudi},
  {Ilyin}, {Johnson}, \& {Strassmeier}}]{2022ApJ...931..111S}
{Stephan}, A.~P., {Wang}, J., {Cauley}, P.~W., {et~al.} 2022, \apj, 931, 111

\bibitem[{{Sudarsky} {et~al.}(2000){Sudarsky}, {Burrows}, \&
  {Pinto}}]{2000ApJ...538..885S}
{Sudarsky}, D., {Burrows}, A., \& {Pinto}, P. 2000, \apj, 538, 885

\bibitem[{{Szab{\'o}} {et~al.}(2021){Szab{\'o}}, {Gandolfi}, {Brandeker},
  {Csizmadia}, {Garai}, {Billot}, {Broeg}, {Ehrenreich}, {Fortier}, {Fossati},
  {Hoyer}, {Kiss}, {Lecavelier des Etangs}, {Maxted}, {Ribas}, {Alibert},
  {Alonso}, {Anglada Escud{\'e}}, {B{\'a}rczy}, {Barros}, {Barrado},
  {Baumjohann}, {Beck}, {Beck}, {Bekkelien}, {Bonfils}, {Benz}, {Borsato},
  {Busch}, {Cabrera}, {Charnoz}, {Collier Cameron}, {Van Damme}, {Davies},
  {Delrez}, {Deleuil}, {Demangeon}, {Demory}, {Erikson}, {Fridlund}, {Futyan},
  {Garc{\'\i}a Mu{\~n}oz}, {Gillon}, {Guedel}, {Guterman}, {Heng}, {Isaak},
  {Lacedelli}, {Laskar}, {Lendl}, {Lovis}, {Luntzer}, {Magrin}, {Nascimbeni},
  {Olofsson}, {Osborn}, {Ottensamer}, {Pagano}, {Pall{\'e}}, {Peter}, {Piazza},
  {Piotto}, {Pollacco}, {Queloz}, {Ragazzoni}, {Rando}, {Rauer}, {Santos},
  {Scandariato}, {S{\'e}gransan}, {Serrano}, {Sicilia}, {Simon}, {Smith},
  {Sousa}, {Steller}, {Thomas}, {Udry}, {Van Grootel}, {Walton}, \&
  {Wilson}}]{2021A&A...654A.159S}
{Szab{\'o}}, G.~M., {Gandolfi}, D., {Brandeker}, A., {et~al.} 2021, \aap, 654,
  A159

\bibitem[{{Szab{\'o}} {et~al.}(2020){Szab{\'o}}, {Pribulla}, {P{\'a}l},
  {B{\'o}di}, {Kiss}, \& {Derekas}}]{2020MNRAS.492L..17S}
{Szab{\'o}}, G.~M., {Pribulla}, T., {P{\'a}l}, A., {et~al.} 2020, \mnras, 492,
  L17

\bibitem[{{Szab{\'o}} {et~al.}(2014){Szab{\'o}}, {Simon}, \&
  {Kiss}}]{2014MNRAS.437.1045S}
{Szab{\'o}}, G.~M., {Simon}, A., \& {Kiss}, L.~L. 2014, \mnras, 437, 1045

\bibitem[{{Temple} {et~al.}(2017){Temple}, {Hellier}, {Albrow}, {Anderson},
  {Bayliss}, {Beatty}, {Bieryla}, {Brown}, {Cargile}, {Collier Cameron},
  {Collins}, {Col{\'o}n}, {Curtis}, {D'Ago}, {Delrez}, {Eastman}, {Gaudi},
  {Gillon}, {Gregorio}, {James}, {Jehin}, {Joner}, {Kielkopf}, {Kuhn},
  {Labadie-Bartz}, {Latham}, {Lendl}, {Lund}, {Malpas}, {Maxted}, {Myers},
  {Oberst}, {Pepe}, {Pepper}, {Pollacco}, {Queloz}, {Rodriguez},
  {S{\'e}gransan}, {Siverd}, {Smalley}, {Stassun}, {Stevens}, {Stockdale},
  {Tan}, {Triaud}, {Udry}, {Villanueva}, {West}, \&
  {Zhou}}]{2017MNRAS.471.2743T}
{Temple}, L.~Y., {Hellier}, C., {Albrow}, M.~D., {et~al.} 2017, \mnras, 471,
  2743

\bibitem[{{Tinetti} {et~al.}(2021){Tinetti}, {Eccleston}, {Haswell}, {Lagage},
  {Leconte}, {L{\"u}ftinger}, {Micela}, {Min}, {Pilbratt}, {Puig}, {Swain},
  {Testi}, {Turrini}, {Vandenbussche}, {Rosa Zapatero Osorio}, {Aret},
  {Beaulieu}, {Buchhave}, {Ferus}, {Griffin}, {Guedel}, {Hartogh}, {Machado},
  {Malaguti}, {Pall{\'e}}, {Rataj}, {Ray}, {Ribas}, {Szab{\'o}}, {Tan},
  {Werner}, {Ratti}, {Scharmberg}, {Salvignol}, {Boudin}, {Halain}, {Haag},
  {Crouzet}, {Kohley}, {Symonds}, {Renk}, {Caldwell}, {Abreu}, {Alonso},
  {Amiaux}, {Berth{\'e}}, {Bishop}, {Bowles}, {Carmona}, {Coffey},
  {Colom{\'e}}, {Crook}, {D{\'e}sjonqueres}, {D{\'\i}az}, {Drummond},
  {Focardi}, {G{\'o}mez}, {Holmes}, {Krijger}, {Kovacs}, {Hunt}, {Machado},
  {Morgante}, {Ollivier}, {Ottensamer}, {Pace}, {Pagano}, {Pascale}, {Pearson},
  {M{\o}ller Pedersen}, {Pniel}, {Roose}, {Savini}, {Stamper}, {Szirovicza},
  {Szoke}, {Tosh}, {Vilardell}, {Barstow}, {Borsato}, {Casewell}, {Changeat},
  {Charnay}, {Civi{\v{s}}}, {Coud{\'e} du Foresto}, {Coustenis}, {Cowan},
  {Danielski}, {Demangeon}, {Drossart}, {Edwards}, {Gilli}, {Encrenaz}, {Kiss},
  {Kokori}, {Ikoma}, {Morales}, {Mendon{\c{c}}a}, {Moneti}, {Mugnai},
  {Garc{\'\i}a Mu{\~n}oz}, {Helled}, {Kama}, {Miguel}, {Nikolaou}, {Pagano},
  {Panic}, {Rengel}, {Rickman}, {Rocchetto}, {Sarkar}, {Selsis}, {Tennyson},
  {Tsiaras}, {Venot}, {Vida}, {Waldmann}, {Yurchenko}, {Szab{\'o}}, {Zellem},
  {Al-Refaie}, {Perez Alvarez}, {Anisman}, {Arhancet}, {Ateca}, {Baeyens},
  {Barnes}, {Bell}, {Benatti}, {Biazzo}, {B{\l}{\k{e}}cka}, {Bonomo}, {Bosch},
  {Bossini}, {Bourgalais}, {Brienza}, {Brucalassi}, {Bruno}, {Caines},
  {Calcutt}, {Campante}, {Canestrari}, {Cann}, {Casali}, {Casas}, {Cassone},
  {Cara}, {Carmona}, {Carone}, {Carrasco}, {Changeat}, {Chioetto},
  {Cortecchia}, {Czupalla}, {Chubb}, {Ciaravella}, {Claret}, {Claudi},
  {Codella}, {Garcia Comas}, {Cracchiolo}, {Cubillos}, {Da Peppo}, {Decin},
  {Dejabrun}, {Delgado-Mena}, {Di Giorgio}, {Diolaiti}, {Dorn}, {Doublier},
  {Doumayrou}, {Dransfield}, {Dumaye}, {Dunford}, {Jimenez Escobar}, {Van
  Eylen}, {Farina}, {Fedele}, {Fern{\'a}ndez}, {Fleury}, {Fonte}, {Fontignie},
  {Fossati}, {Funke}, {Galy}, {Garai}, {Garc{\'\i}a}, {Garc{\'\i}a-Rigo},
  {Garufi}, {Germano Sacco}, {Giacobbe}, {G{\'o}mez}, {Gonzalez},
  {Gonzalez-Galindo}, {Grassi}, {Griffith}, {Guarcello}, {Goujon}, {Gressier},
  {Grzegorczyk}, {Guillot}, {Guilluy}, {Hargrave}, {Hellin}, {Herrero},
  {Hills}, {Horeau}, {Ito}, {Jessen}, {Kabath}, {K{\'a}lm{\'a}n}, {Kawashima},
  {Kimura}, {Kn{\'\i}{\v{z}}ek}, {Kreidberg}, {Kruid}, {Kruijssen},
  {Kubel{\'\i}k}, {Lara}, {Lebonnois}, {Lee}, {Lefevre}, {Lichtenberg},
  {Locci}, {Lombini}, {Sanchez Lopez}, {Lorenzani}, {MacDonald}, {Magrini},
  {Maldonado}, {Marcq}, {Migliorini}, {Modirrousta-Galian}, {Molaverdikhani},
  {Molinari}, {Molli{\`e}re}, {Moreau}, {Morello}, {Morinaud}, {Morvan},
  {Moses}, {Mouzali}, {Nakhjiri}, {Naponiello}, {Narita}, {Nascimbeni},
  {Nikolaou}, {Noce}, {Oliva}, {Palladino}, {Papageorgiou}, {Parmentier},
  {Peres}, {P{\'e}rez}, {Perez-Hoyos}, {Perger}, {Cecchi Pestellini},
  {Petralia}, {Philippon}, {Piccialli}, {Pignatari}, {Piotto}, {Podio},
  {Polenta}, {Preti}, {Pribulla}, {Lopez Puertas}, {Rainer}, {Reess}, {Rimmer},
  {Robert}, {Rosich}, {Rossi}, {Rust}, {Saleh}, {Sanna}, {Schisano},
  {Schreiber}, {Schwartz}, {Scippa}, {Seli}, {Shibata}, {Simpson}, {Shorttle},
  {Skaf}, {Skup}, {Sobiecki}, {Sousa}, {Sozzetti}, {{\v{S}}poner}, {Steiger},
  {Tanga}, {Tackley}, {Taylor}, {Tecza}, {Terenzi}, {Tremblin}, {Tozzi},
  {Triaud}, {Trompet}, {Tsai}, {Tsantaki}, {Valencia}, {Carine Vandaele}, {Van
  der Swaelmen}, {Adibekyan}, {Vasisht}, {Vazan}, {Del Vecchio}, {Waltham},
  {Wawer}, {Widemann}, {Wolkenberg}, {Hou Yip}, {Yung}, {Zilinskas},
  {Zingales}, \& {Zuppella}}]{2021arXiv210404824T}
{Tinetti}, G., {Eccleston}, P., {Haswell}, C., {et~al.} 2021, arXiv e-prints,
  arXiv:2104.04824

\bibitem[{{von Essen} {et~al.}(2014){von Essen}, {Czesla}, {Wolter}, {Breger},
  {Herrero}, {Mallonn}, {Ribas}, {Strassmeier}, \&
  {Morales}}]{2014A&A...561A..48V}
{von Essen}, C., {Czesla}, S., {Wolter}, U., {et~al.} 2014, \aap, 561, A48

\bibitem[{{von Essen} {et~al.}(2015){von Essen}, {Mallonn}, {Albrecht},
  {Antoci}, {Smith}, {Dreizler}, \& {Strassmeier}}]{2015A&A...584A..75V}
{von Essen}, C., {Mallonn}, M., {Albrecht}, S., {et~al.} 2015, \aap, 584, A75

\bibitem[{{von Essen} {et~al.}(2020){von Essen}, {Mallonn}, {Borre}, {Antoci},
  {Stassun}, {Khalafinejad}, \&
  {Tautvai{\v{s}}ien{\.{e}}}}]{2020A&A...639A..34V}
{von Essen}, C., {Mallonn}, M., {Borre}, C.~C., {et~al.} 2020, \aap, 639, A34

\bibitem[{{Watanabe} {et~al.}(2022){Watanabe}, {Narita}, {Palle}, {Fukui},
  {Kusakabe}, {Parviainen}, {Murgas}, {Casasayas-Barris}, {Johnson}, {Sato},
  {Livingston}, {de Leon}, {Mori}, {Nishiumi}, {Terada}, {Esparza-Borges}, \&
  {Kawauchi}}]{2022MNRAS.512.4404W}
{Watanabe}, N., {Narita}, N., {Palle}, E., {et~al.} 2022, \mnras, 512, 4404

\bibitem[{{Wong} {et~al.}(2020){Wong}, {Shporer}, {Daylan}, {Benneke},
  {Fetherolf}, {Kane}, {Ricker}, {Vanderspek}, {Latham}, {Winn}, {Jenkins},
  {Boyd}, {Glidden}, {Goeke}, {Sha}, {Ting}, \&
  {Yahalomi}}]{2020AJ....160..155W}
{Wong}, I., {Shporer}, A., {Daylan}, T., {et~al.} 2020, \aj, 160, 155

\bibitem[{{Zhang} {et~al.}(2018){Zhang}, {Knutson}, {Kataria}, {Schwartz},
  {Cowan}, {Showman}, {Burrows}, {Fortney}, {Todorov}, {Desert}, {Agol}, \&
  {Deming}}]{2018AJ....155...83Z}
{Zhang}, M., {Knutson}, H.~A., {Kataria}, T., {et~al.} 2018, \aj, 155, 83

\bibitem[{{Zucker} {et~al.}(2007){Zucker}, {Mazeh}, \&
  {Alexander}}]{2007ApJ...670.1326Z}
{Zucker}, S., {Mazeh}, T., \& {Alexander}, T. 2007, \apj, 670, 1326

\end{thebibliography}

\begin{appendix}
    \section{Influence of the stellar pulsations on the out-of-transit parameters}



\begin{table*}
\caption{Behaviour of the monitored out-of-transit parameters and $R_{\rm p}/R_S$ with respect to the number of subsequent sinusoidal curves (with decreasing amplitudes) used in the whitening process.}
\label{tab:oot_params}
\centering
\scriptsize
\begin{tabular}{l c c c c c c c c}
\hline
\hline
 & $K_{\rm phot}$ [m/s] & $q_{\rm{ell}}$ & $A_g$ & $I_{\rm p}/I_\star$ & $\varepsilon$ [$^\circ$] & Occ. depth [ppm] & $R_{\rm p}/R_\star$ & BIC \\
\hline
$\Phi_{0}$ & $2205.5 \pm 603.8$ & $0.01187 \pm 0.00138$ & $0.72 \pm 0.16$ & $0.0126 \pm 0.0079$ & $-83.1 \pm 11.8$ & $191.1 \pm 27.0$ & $0.09054 \pm 0.00053$ & $-594783.8$ \\ 
$\Phi_{1}$ & $2576.1 \pm 662.5$ & $0.01362 \pm 0.00162$ & $0.91 \pm 0.18$ & $0.0101 \pm 0.0076$ & $-73.2 \pm 9.2$ & $251.1 \pm 26.0$ & $0.09064 \pm 0.00061$ & $-578129.8$ \\ 
$\Phi_{2}$ & $82.1 \pm 56.7$ & $0.00040 \pm 0.00024$ & $0.88 \pm 0.19$ & $0.0042 \pm 0.0045$ & $-146.6 \pm 7.1$ & $18.4 \pm 25.9$ & $0.09036 \pm 0.00061$ & $-579458.4$ \\ 
$\Phi_{3}$ & $77.2 \pm 60.0$ & $0.00040 \pm 0.00026$ & $0.42 \pm 0.14$ & $0.0022 \pm 0.0030$ & $-12.8 \pm 15.7$ & $167.5 \pm 25.9$ & $0.09013 \pm 0.00059$ & $-579477.8$ \\ 
$\Phi_{4}$ & $75.2 \pm 56.8$ & $0.00037 \pm 0.00024$ & $0.43 \pm 0.13$ & $0.0022 \pm 0.0027$ & $-7.7 \pm 14.8$ & $171.4 \pm 25.9$ & $0.09007 \pm 0.00055$ & $-579858.6$ \\ 
$\Phi_{5}$ & $83.7 \pm 64.1$ & $0.00043 \pm 0.00027$ & $0.39 \pm 0.15$ & $0.0024 \pm 0.0031$ & $3.3 \pm 15.3$ & $157.1 \pm 25.9$ & $0.08999 \pm 0.00064$ & $-579584.7$ \\ 
$\Phi_{6}$ & $64.2 \pm 49.2$ & $0.00032 \pm 0.00022$ & $0.40 \pm 0.13$ & $0.0014 \pm 0.0018$ & $7.1 \pm 12.6$ & $135.5 \pm 25.9$ & $0.08951 \pm 0.00059$ & $-579963.9$ \\ 
$\Phi_{7}$ & $65.7 \pm 50.9$ & $0.00033 \pm 0.00020$ & $0.40 \pm 0.13$ & $0.0013 \pm 0.0019$ & $-6.6 \pm 16.0$ & $132.9 \pm 25.9$ & $0.08952 \pm 0.00059$ & $-581147.2$ \\ 
$\Phi_{8}$ & $66.9 \pm 48.1$ & $0.00033 \pm 0.00021$ & $0.37 \pm 0.14$ & $0.0015 \pm 0.0018$ & $-7.7 \pm 15.8$ & $130.5 \pm 26.0$ & $0.08971 \pm 0.00061$ & $-581641.9$ \\ 
$\Phi_{9}$ & $65.5 \pm 49.3$ & $0.00032 \pm 0.00020$ & $0.34 \pm 0.14$ & $0.0015 \pm 0.0018$ & $-4.5 \pm 19.6$ & $119.1 \pm 26.0$ & $0.08967 \pm 0.00059$ & $-581367.6$ \\ 
$\Phi_{10}$ & $65.8 \pm 48.9$ & $0.00032 \pm 0.00020$ & $0.33 \pm 0.15$ & $0.0016 \pm 0.0020$ & $-4.5 \pm 18.3$ & $119.6 \pm 26.0$ & $0.08977 \pm 0.00059$ & $-582065.9$ \\ 
$\Phi_{11}$ & $67.3 \pm 48.3$ & $0.00033 \pm 0.00021$ & $0.39 \pm 0.14$ & $0.0019 \pm 0.0024$ & $-13.4 \pm 19.1$ & $143.0 \pm 26.1$ & $0.08989 \pm 0.00058$ & $-582802.5$ \\ 
$\Phi_{12}$ & $64.5 \pm 47.5$ & $0.00032 \pm 0.00020$ & $0.38 \pm 0.13$ & $0.0016 \pm 0.0021$ & $2.9 \pm 13.5$ & $147.8 \pm 26.2$ & $0.08992 \pm 0.00059$ & $-586007.5$ \\ 
$\Phi_{13}$ & $68.3 \pm 53.2$ & $0.00034 \pm 0.00020$ & $0.30 \pm 0.14$ & $0.0019 \pm 0.0024$ & $7.1 \pm 19.1$ & $105.0 \pm 26.2$ & $0.08963 \pm 0.00058$ & $-586096.1$ \\ 
$\Phi_{14}$ & $69.5 \pm 50.9$ & $0.00034 \pm 0.00022$ & $0.30 \pm 0.14$ & $0.0022 \pm 0.0026$ & $3.1 \pm 21.4$ & $116.2 \pm 26.3$ & $0.08965 \pm 0.00058$ & $-587206.1$ \\ 
$\Phi_{15}$ & $46.9 \pm 35.1$ & $0.00023 \pm 0.00014$ & $0.39 \pm 0.11$ & $0.0009 \pm 0.0012$ & $20.7 \pm 10.3$ & $122.8 \pm 27.3$ & $0.08934 \pm 0.00053$ & $-600450.9$ \\ 
$\Phi_{16}$ & $45.3 \pm 33.6$ & $0.00022 \pm 0.00014$ & $0.34 \pm 0.11$ & $0.0010 \pm 0.0013$ & $17.7 \pm 11.2$ & $106.8 \pm 27.3$ & $0.08945 \pm 0.00051$ & $-600913.6$ \\ 
\hline
\end{tabular}
\end{table*}

In order to test how the stellar pulsations affect the precision and accuracy of out-of-transit parameters, we prepared 17 LCs according to Eq. (\ref{eq:prep}). The monitored parameters include the amplitude of the Doppler beaming ($K_{\rm phot}$), the ellipsoidal mass ratio ($q_{\rm ell}$), the geometric albedo ($A_g$), the surface brightness ratio ($I_{\rm p}/I_\star$), the offset between the brightest point on the dayside hemisphere and the substellar point ($\varepsilon$) and the occultation depth. The behaviour of the relative planetary radius is also observed for consistency. These are listed in Table \ref{tab:oot_params}. In order to distinguish between the 17 sets of solutions, we utilize the Bayesian Information Criterion (BIC), defined as:
\begin{equation}
    \text{BIC} = n \ln \left( \frac{\text{RSS}}{n} \right) + k \ln n,
\end{equation}
where $n$ is the number of LC points and $k$ is number of parameters ($13$ plus three times the number of sinusiodial curves subtracted). We establish that the LC solutions computed from $\Phi_{16}(t)$ have the lowest BIC value, and accept this set of parameters.

\section{Consistency check}
\begin{figure}
    \centering
    \includegraphics[width = \columnwidth]{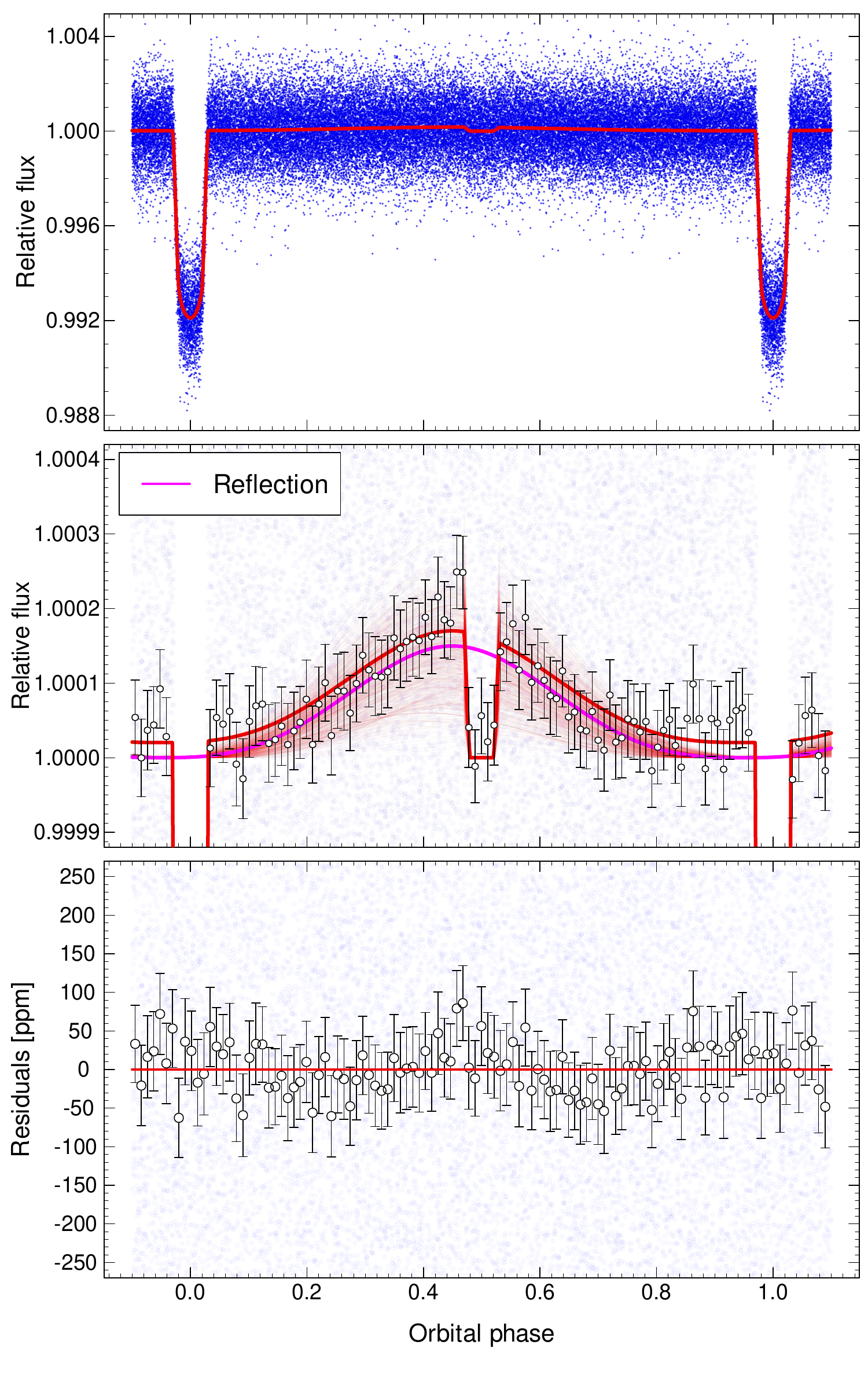}
    \caption{Same as Fig. \ref{fig:occ} but with $0$ mass assumed for WASP-167b.}
    \label{fig:occ_nomass}
\end{figure}

We performed a check for consistency for the phase curve analysis by excluding the Doppler beaming and the ellipsoidal variability from the modelling. The resultant LC model is shown in Fig. \ref{fig:occ_nomass}. 

We then performed a second consistency check, where we modelled the data from the three sectors independently from each other, from $\Phi_0$ to $\Phi_{16}$. We monitored the behaviour of the same parameters as in Figs. \ref{fig:Kq} and \ref{fig:consts}.

\begin{figure*}
    \centering
    \includegraphics[width = \textwidth]{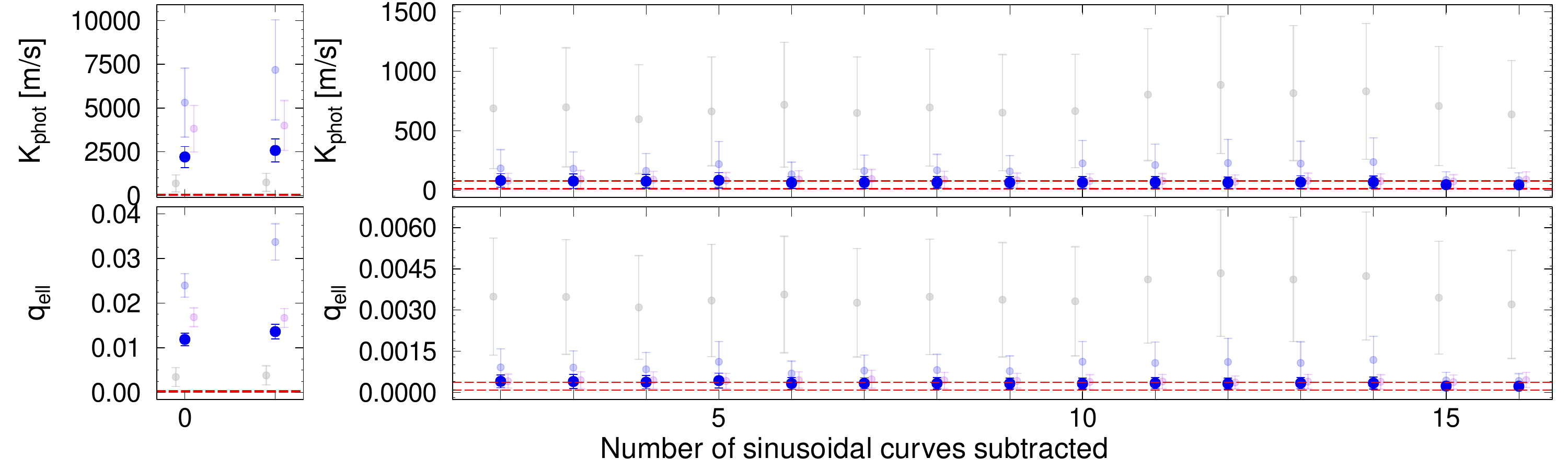}
    \caption{Behaviour of the best-fit $K_{\rm phot}$ and $q_{\rm ell}$ parameters with respect to the pre-whitening. The dark blue dots and error bars, as well as the red dashed lines, are the same as on Fig. \ref{fig:Kq}. The respective parameters from Sector 10, 37 and 64 are shown with grey, light blue and pink, respectively. The latter points and error bars are also slightly offset on the x axis for better visibility.}
    \label{fig:kq_sep_sec}
\end{figure*}

\begin{figure*}
    \centering
    \includegraphics[width = \textwidth]{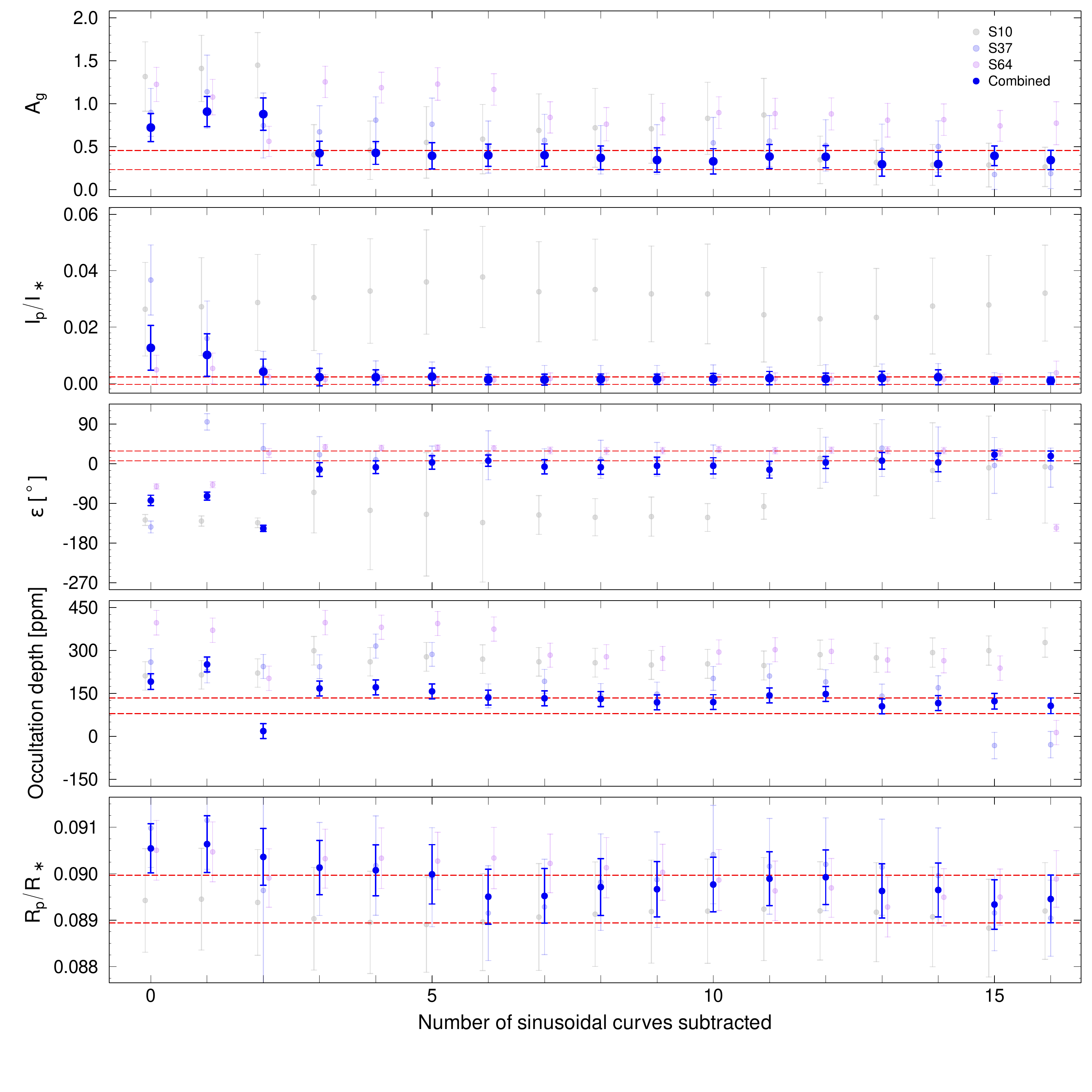}
    \caption{Behaviour of the best-fit $A_g$, $I_p/I_\star$, occultation depth and $R_{\rm p}/R_\star$ parameters with respect to the pre-whitening. The dark blue dots and error bars, as well as the red dashed lines, are the same as on Fig. \ref{fig:consts}. The respective parameters from Sector 10, 37 and 64 are shown with grey, light blue and pink, respectively. The latter points and error bars are also slightly offset on the x axis for better visibility.}
    \label{fig:consts_sep_sec}
\end{figure*}

\begin{table}
\caption{System parameters from Sectors 10, 37 and 64.}
\label{tab:params_comp_sep_sec}
\centering
\scriptsize
\begin{tabular}{l c c c c c c c}
\hline
\hline
 & $K$ [m/s] & $q_{\rm{ell}}$ & $A_g$ & $I_{\rm p}/I_\star$ & $\varepsilon$ [$^\circ$] & Occ. depth [ppm] & $R_{\rm p}/R_\star$ \\
\hline
\multicolumn{8}{c}{Sector 10} \\
\hline
$\Phi_{0}$ & $691.2 \pm 489.7$ & $0.00344 \pm 0.00209$ & $1.32 \pm 0.40$ & $0.0263 \pm 0.0166$ & $-127.4 \pm 12.3$ & $211.3 \pm 49.5$ & $0.08942 \pm 0.00111$ \\ 
$\Phi_{1}$ & $749.9 \pm 511.7$ & $0.00381 \pm 0.00214$ & $1.41 \pm 0.38$ & $0.0272 \pm 0.0174$ & $-129.9 \pm 11.5$ & $214.6 \pm 49.5$ & $0.08945 \pm 0.00110$ \\ 
$\Phi_{2}$ & $688.6 \pm 507.4$ & $0.00349 \pm 0.00213$ & $1.45 \pm 0.38$ & $0.0287 \pm 0.0170$ & $-133.7 \pm 11.1$ & $221.0 \pm 49.4$ & $0.08938 \pm 0.00114$ \\ 
$\Phi_{3}$ & $697.6 \pm 500.3$ & $0.00348 \pm 0.00209$ & $0.41 \pm 0.35$ & $0.0305 \pm 0.0188$ & $-64.8 \pm 92.1$ & $299.0 \pm 49.8$ & $0.08903 \pm 0.00110$ \\ 
$\Phi_{4}$ & $597.3 \pm 456.9$ & $0.00309 \pm 0.00189$ & $0.46 \pm 0.34$ & $0.0328 \pm 0.0186$ & $-105.6 \pm 134.6$ & $260.4 \pm 49.9$ & $0.08895 \pm 0.00110$ \\ 
$\Phi_{5}$ & $662.9 \pm 456.8$ & $0.00335 \pm 0.00204$ & $0.55 \pm 0.42$ & $0.0360 \pm 0.0185$ & $-114.7 \pm 139.9$ & $277.8 \pm 50.0$ & $0.08891 \pm 0.00103$ \\ 
$\Phi_{6}$ & $718.6 \pm 525.2$ & $0.00356 \pm 0.00212$ & $0.59 \pm 0.40$ & $0.0378 \pm 0.0179$ & $-133.3 \pm 135.1$ & $269.7 \pm 50.1$ & $0.08896 \pm 0.00104$ \\ 
$\Phi_{7}$ & $649.8 \pm 472.0$ & $0.00326 \pm 0.00198$ & $0.69 \pm 0.43$ & $0.0325 \pm 0.0178$ & $-116.1 \pm 44.1$ & $260.4 \pm 50.0$ & $0.08907 \pm 0.00115$ \\ 
$\Phi_{8}$ & $696.1 \pm 491.5$ & $0.00348 \pm 0.00211$ & $0.72 \pm 0.46$ & $0.0333 \pm 0.0179$ & $-121.3 \pm 42.2$ & $257.0 \pm 50.1$ & $0.08913 \pm 0.00113$ \\ 
$\Phi_{9}$ & $652.4 \pm 489.0$ & $0.00337 \pm 0.00208$ & $0.71 \pm 0.40$ & $0.0318 \pm 0.0170$ & $-119.8 \pm 44.3$ & $249.2 \pm 50.3$ & $0.08918 \pm 0.00111$ \\ 
$\Phi_{10}$ & $665.6 \pm 476.5$ & $0.00332 \pm 0.00199$ & $0.83 \pm 0.42$ & $0.0318 \pm 0.0177$ & $-121.7 \pm 31.5$ & $253.1 \pm 50.4$ & $0.08919 \pm 0.00112$ \\ 
$\Phi_{11}$ & $803.8 \pm 554.8$ & $0.00411 \pm 0.00232$ & $0.87 \pm 0.43$ & $0.0244 \pm 0.0168$ & $-96.8 \pm 29.1$ & $247.3 \pm 50.6$ & $0.08924 \pm 0.00111$ \\ 
$\Phi_{12}$ & $885.9 \pm 575.8$ & $0.00434 \pm 0.00230$ & $0.35 \pm 0.28$ & $0.0229 \pm 0.0165$ & $12.5 \pm 68.0$ & $285.3 \pm 50.7$ & $0.08920 \pm 0.00106$ \\ 
$\Phi_{13}$ & $816.2 \pm 566.9$ & $0.00411 \pm 0.00226$ & $0.32 \pm 0.26$ & $0.0234 \pm 0.0174$ & $9.4 \pm 81.8$ & $274.5 \pm 50.9$ & $0.08917 \pm 0.00107$ \\ 
$\Phi_{14}$ & $831.8 \pm 570.1$ & $0.00424 \pm 0.00233$ & $0.29 \pm 0.24$ & $0.0274 \pm 0.0170$ & $-15.3 \pm 108.4$ & $292.5 \pm 51.0$ & $0.08907 \pm 0.00107$ \\ 
$\Phi_{15}$ & $708.6 \pm 500.8$ & $0.00345 \pm 0.00205$ & $0.29 \pm 0.25$ & $0.0279 \pm 0.0175$ & $-9.1 \pm 117.3$ & $299.5 \pm 51.1$ & $0.08883 \pm 0.00105$ \\ 
$\Phi_{16}$ & $638.0 \pm 452.4$ & $0.00321 \pm 0.00197$ & $0.26 \pm 0.23$ & $0.0320 \pm 0.0170$ & $-6.7 \pm 128.0$ & $327.5 \pm 51.3$ & $0.08920 \pm 0.00104$ \\ 
\hline
\multicolumn{8}{c}{Sector 37} \\
\hline
$\Phi_{0}$ & $5310.9 \pm 1970.7$ & $0.02396 \pm 0.00265$ & $0.90 \pm 0.28$ & $0.0367 \pm 0.0124$ & $-143.4 \pm 13.5$ & $259.3 \pm 46.7$ & $0.09098 \pm 0.00085$ \\ 
$\Phi_{1}$ & $7180.8 \pm 2857.6$ & $0.03372 \pm 0.00409$ & $1.14 \pm 0.43$ & $0.0160 \pm 0.0133$ & $95.2 \pm 18.7$ & $228.5 \pm 41.9$ & $0.09115 \pm 0.00105$ \\ 
$\Phi_{2}$ & $183.7 \pm 157.7$ & $0.00091 \pm 0.00067$ & $0.75 \pm 0.38$ & $0.0055 \pm 0.0060$ & $34.4 \pm 56.8$ & $243.8 \pm 41.7$ & $0.08964 \pm 0.00337$ \\ 
$\Phi_{3}$ & $181.4 \pm 140.7$ & $0.00090 \pm 0.00061$ & $0.67 \pm 0.30$ & $0.0049 \pm 0.0056$ & $20.6 \pm 41.1$ & $243.1 \pm 41.7$ & $0.09010 \pm 0.00100$ \\ 
$\Phi_{4}$ & $163.3 \pm 144.9$ & $0.00084 \pm 0.00062$ & $0.81 \pm 0.27$ & $0.0037 \pm 0.0043$ & $9.4 \pm 18.6$ & $315.4 \pm 41.8$ & $0.09018 \pm 0.00107$ \\ 
$\Phi_{5}$ & $219.4 \pm 189.4$ & $0.00111 \pm 0.00073$ & $0.76 \pm 0.30$ & $0.0034 \pm 0.0042$ & $17.7 \pm 22.7$ & $286.3 \pm 41.5$ & $0.08993 \pm 0.00107$ \\ 
$\Phi_{6}$ & $133.6 \pm 103.4$ & $0.00069 \pm 0.00045$ & $0.50 \pm 0.30$ & $0.0027 \pm 0.0032$ & $19.4 \pm 55.7$ & $141.4 \pm 41.5$ & $0.08915 \pm 0.00102$ \\ 
$\Phi_{7}$ & $162.3 \pm 134.4$ & $0.00079 \pm 0.00056$ & $0.57 \pm 0.30$ & $0.0029 \pm 0.0035$ & $5.8 \pm 28.3$ & $192.3 \pm 41.5$ & $0.08929 \pm 0.00103$ \\ 
$\Phi_{8}$ & $167.9 \pm 134.4$ & $0.00081 \pm 0.00057$ & $0.46 \pm 0.28$ & $0.0028 \pm 0.0036$ & $10.2 \pm 43.1$ & $142.5 \pm 41.5$ & $0.08982 \pm 0.00104$ \\ 
$\Phi_{9}$ & $158.6 \pm 132.7$ & $0.00077 \pm 0.00055$ & $0.47 \pm 0.29$ & $0.0029 \pm 0.0035$ & $9.9 \pm 38.5$ & $147.8 \pm 41.5$ & $0.08987 \pm 0.00103$ \\ 
$\Phi_{10}$ & $226.1 \pm 193.4$ & $0.00112 \pm 0.00073$ & $0.54 \pm 0.30$ & $0.0031 \pm 0.0035$ & $5.0 \pm 37.8$ & $202.3 \pm 41.6$ & $0.09041 \pm 0.00106$ \\ 
$\Phi_{11}$ & $212.6 \pm 174.4$ & $0.00107 \pm 0.00076$ & $0.57 \pm 0.30$ & $0.0027 \pm 0.0031$ & $1.8 \pm 27.7$ & $210.7 \pm 41.9$ & $0.09016 \pm 0.00103$ \\ 
$\Phi_{12}$ & $228.8 \pm 198.3$ & $0.00111 \pm 0.00086$ & $0.52 \pm 0.30$ & $0.0031 \pm 0.0036$ & $9.3 \pm 51.1$ & $190.3 \pm 42.0$ & $0.09020 \pm 0.00100$ \\ 
$\Phi_{13}$ & $224.2 \pm 188.0$ & $0.00107 \pm 0.00077$ & $0.46 \pm 0.30$ & $0.0031 \pm 0.0038$ & $35.5 \pm 64.4$ & $140.5 \pm 42.1$ & $0.09014 \pm 0.00103$ \\ 
$\Phi_{14}$ & $237.2 \pm 202.7$ & $0.00118 \pm 0.00086$ & $0.50 \pm 0.30$ & $0.0033 \pm 0.0037$ & $21.6 \pm 62.1$ & $169.8 \pm 42.0$ & $0.08996 \pm 0.00103$ \\ 
$\Phi_{15}$ & $86.5 \pm 67.6$ & $0.00043 \pm 0.00030$ & $0.18 \pm 0.17$ & $0.0017 \pm 0.0021$ & $-3.9 \pm 63.4$ & $-31.9 \pm 46.1$ & $0.08916 \pm 0.00082$ \\ 
$\Phi_{16}$ & $85.3 \pm 62.0$ & $0.00041 \pm 0.00027$ & $0.19 \pm 0.17$ & $0.0017 \pm 0.0021$ & $-8.6 \pm 44.5$ & $-29.1 \pm 46.1$ & $0.08904 \pm 0.00082$ \\ 
\hline
\multicolumn{8}{c}{Sector 64} \\
\hline
$\Phi_{0}$ & $3821.1 \pm 1332.4$ & $0.01684 \pm 0.00210$ & $1.23 \pm 0.20$ & $0.0048 \pm 0.0052$ & $-51.2 \pm 6.0$ & $396.6 \pm 42.7$ & $0.09050 \pm 0.00064$ \\ 
$\Phi_{1}$ & $4002.0 \pm 1426.3$ & $0.01670 \pm 0.00211$ & $1.08 \pm 0.21$ & $0.0053 \pm 0.0055$ & $-47.5 \pm 6.8$ & $370.5 \pm 42.6$ & $0.09047 \pm 0.00064$ \\ 
$\Phi_{2}$ & $79.6 \pm 62.0$ & $0.00041 \pm 0.00025$ & $0.56 \pm 0.18$ & $0.0024 \pm 0.0026$ & $24.7 \pm 10.2$ & $202.4 \pm 42.5$ & $0.08991 \pm 0.00063$ \\ 
$\Phi_{3}$ & $91.4 \pm 74.3$ & $0.00044 \pm 0.00031$ & $1.25 \pm 0.18$ & $0.0014 \pm 0.0017$ & $38.1 \pm 5.9$ & $397.1 \pm 42.2$ & $0.09032 \pm 0.00064$ \\ 
$\Phi_{4}$ & $89.4 \pm 68.9$ & $0.00044 \pm 0.00030$ & $1.19 \pm 0.18$ & $0.0014 \pm 0.0016$ & $36.0 \pm 6.4$ & $380.8 \pm 42.1$ & $0.09033 \pm 0.00065$ \\ 
$\Phi_{5}$ & $83.0 \pm 66.1$ & $0.00042 \pm 0.00028$ & $1.23 \pm 0.19$ & $0.0011 \pm 0.0015$ & $36.8 \pm 5.8$ & $394.1 \pm 42.1$ & $0.09027 \pm 0.00062$ \\ 
$\Phi_{6}$ & $90.7 \pm 73.6$ & $0.00044 \pm 0.00030$ & $1.17 \pm 0.18$ & $0.0012 \pm 0.0016$ & $35.4 \pm 6.2$ & $374.4 \pm 42.1$ & $0.09034 \pm 0.00066$ \\ 
$\Phi_{7}$ & $95.0 \pm 79.2$ & $0.00048 \pm 0.00033$ & $0.84 \pm 0.18$ & $0.0017 \pm 0.0020$ & $30.2 \pm 7.4$ & $283.6 \pm 42.1$ & $0.09022 \pm 0.00063$ \\ 
$\Phi_{8}$ & $90.2 \pm 68.4$ & $0.00044 \pm 0.00029$ & $0.76 \pm 0.19$ & $0.0016 \pm 0.0020$ & $29.0 \pm 8.3$ & $278.0 \pm 42.1$ & $0.09013 \pm 0.00065$ \\ 
$\Phi_{9}$ & $83.5 \pm 60.9$ & $0.00042 \pm 0.00027$ & $0.82 \pm 0.18$ & $0.0018 \pm 0.0020$ & $30.0 \pm 7.5$ & $272.2 \pm 42.0$ & $0.09003 \pm 0.00060$ \\ 
$\Phi_{10}$ & $77.9 \pm 60.7$ & $0.00038 \pm 0.00026$ & $0.90 \pm 0.19$ & $0.0015 \pm 0.0017$ & $32.2 \pm 7.0$ & $294.5 \pm 42.1$ & $0.08986 \pm 0.00065$ \\ 
$\Phi_{11}$ & $78.7 \pm 60.8$ & $0.00040 \pm 0.00026$ & $0.89 \pm 0.18$ & $0.0017 \pm 0.0020$ & $29.7 \pm 7.3$ & $302.6 \pm 42.2$ & $0.08963 \pm 0.00064$ \\ 
$\Phi_{12}$ & $71.5 \pm 57.0$ & $0.00036 \pm 0.00023$ & $0.88 \pm 0.19$ & $0.0015 \pm 0.0021$ & $30.6 \pm 7.2$ & $296.8 \pm 42.4$ & $0.08969 \pm 0.00063$ \\ 
$\Phi_{13}$ & $80.4 \pm 64.0$ & $0.00039 \pm 0.00026$ & $0.81 \pm 0.20$ & $0.0015 \pm 0.0020$ & $30.7 \pm 7.8$ & $266.7 \pm 42.4$ & $0.08928 \pm 0.00065$ \\ 
$\Phi_{14}$ & $76.8 \pm 61.7$ & $0.00039 \pm 0.00026$ & $0.81 \pm 0.18$ & $0.0015 \pm 0.0018$ & $29.9 \pm 7.3$ & $264.1 \pm 42.5$ & $0.08949 \pm 0.00061$ \\ 
$\Phi_{15}$ & $73.9 \pm 57.4$ & $0.00037 \pm 0.00026$ & $0.74 \pm 0.18$ & $0.0015 \pm 0.0019$ & $26.0 \pm 8.2$ & $238.2 \pm 42.4$ & $0.08950 \pm 0.00060$ \\ 
$\Phi_{16}$ & $89.8 \pm 66.1$ & $0.00045 \pm 0.00028$ & $0.77 \pm 0.25$ & $0.0038 \pm 0.0041$ & $-145.4 \pm 8.0$ & $13.4 \pm 42.5$ & $0.08989 \pm 0.00061$ \\ 
\hline
\end{tabular}
\end{table}

\clearpage

\section{Posterior distribution}

\begin{figure*}
    \centering
    \includegraphics[width = \textwidth]{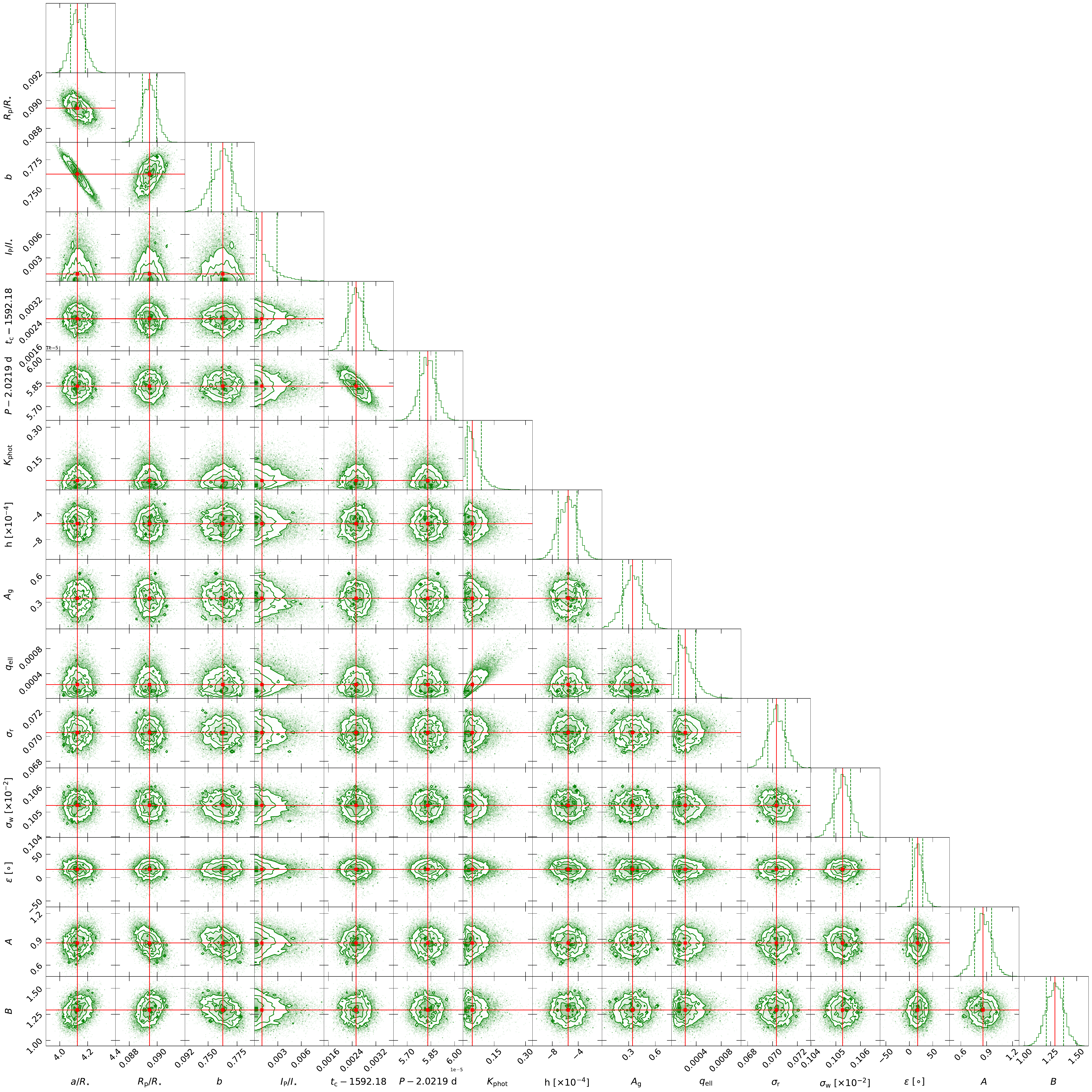}
    \caption{Corner plot showing the posterior distribution from the $\Phi_{16}$ light curve modelling.}
    \label{fig:corner}
\end{figure*}

The posterior distribution from the Markov chain Monte Carlo light curve sampling, prepared via the routines of \cite{2017AJ....154..220F}, are shown in Fig. \ref{fig:corner}.

\end{appendix}

\end{document}